\documentclass[Journal,letterpaper]{ascelike}

\usepackage[utf8]{inputenc}
\usepackage[T1]{fontenc}
\usepackage{lmodern}
\usepackage{graphicx}
\usepackage{amsmath}
\usepackage{lineno}
\usepackage{newtxtext,newtxmath}
\NameTag{Lu, \today}
\begin{document}

\title{Combining plane wave expansion and variational techniques for fast phononic computations}

\author{Yan Lu$^*$ 
\thanks{Research assistant. Department of Department of Mechanical, Materials, and Aerospace Engineering, Illinois Institute of Technology, 10 W. 32nd St Chicago, IL, 60616
USA. $^*$Corresponding author: Email: ylu50@hawk.iit.edu}
 and 
Ankit Srivastava
\thanks{Ph.D., Assistant Professor. Department of Department of Mechanical, Materials, and Aerospace Engineering, Illinois Institute of Technology, 10 W. 32nd St Chicago, IL, 60616
USA. Email: asriva13@iit.edu}
}

\maketitle

\begin{abstract}
In this paper the salient features of the Plane Wave Expansion (PWE) method and the mixed variational technique are combined for the fast eigenvalue computations of arbitrarily complex phononic unit cells. This is done by expanding the material properties in a Fourier expansion, as is the case with PWE. The required matrix elements in the variational scheme are identified as the discrete Fourier transform coefficients of material properties, thus obviating the need for any explicit integration. The process allows us to provide succinct and closed form expressions for all the matrices involved in the mixed variational method. The scheme proposed here preserves both the simplicity of expression which is inherent in the PWE method and the superior convergence properties of the mixed variational scheme. We present numerical results and comment upon the convergence and stability of the current method. We show that the current representation renders the results of the method stable over the entire range of the expansion terms as allowed by the spatial discretization. When compared with a zero order numerical integration scheme, the present method results in greater computational accuracy of all eigenvalues. A higher order numerical integration scheme comes close to the accuracy of the present method but only with significantly more computational expense. 
\end{abstract}


\maketitle

\section{Introduction}
In recent years phononic crystals have emerged as an exciting medium for the fine-tuned control of acoustic and stress waves. They provide potential for applications in such areas as exotic refraction  \cite{cervera2001refractive,sukhovich2008negative,srivastava2016metamaterial,willis2016negative}, ultrasound tunneling \cite{yang2002ultrasound}, waveguiding \cite{khelif2003trapping}, sound focusing \cite{yang2004focusing}, acoustic rectification \cite{li2011tunable}, seismic and shock wave mitigation \cite{bao2011dynamic,mitchell2016effect,dertimanis2016feasibility}, thermal property tuning \cite{cleland2001thermal,landry2008complex,zen2014engineering} and flow stabalization\cite{hussein2015flow}. In addition to these applications certain research areas, such as phononic bandgap optimization \cite{bilal2011ultrawide,lu20173}, topological phononic crystals \cite{chaunsali2016stress} and dynamic homogenization \cite{nemat2011homogenization,nemat2011overall,srivastava2012overall,srivastava2015elastic}, have also raised broad interest in the community. The recent surge of research effort toward the study of wave propagation in phononic crystals has depended upon the speed, efficiency and versatility of phononic bandstructure calculating algorithms. Several techniques can be used to calculate these bandstructures including the plane wave expansion (PWE) method \cite{sigalas1993band,kushwaha1993acoustic,kushwaha1994theory,vasseur1994complete}, the multiple scattering method \cite{kafesaki1999multiple,amirkulova2015acoustic}, the finite difference time domain method \cite{sigalas2000theoretical,cao2004finite,hsieh2006three}, the finite element (FE) method \cite{hladky1991analysis,veres2012complexity} (See \citeNP{hussein2014dynamics}). 

PWE and variational methods are two of the most commonly used solvers owing to the ease of their implementation and their versatility. The salient feature of PWE is that it expresses material properties and the displacement field using Fourier expansion. The expansion terms can be made to satisfy Bloch-periodicity a-priori and PWE is algorithmically easy to implement. However, PWE converges slowly when material properties show large contrast with convergence properties similar to a simple displacement based variational scheme (Rayleigh quotient, see \citeNP{lu2016variational}). FE method is based on variational theories and it is a preferred method for evaluating phononic band structure of various geometries. \citeN{haque2016spatial} have shown that special care should be taken in interpreting the FE results due to spatial aliasing. 

In this paper the salient aspects of PWE and a variational technique, which is based upon varying both the displacement and the stress fields \cite{minagawa1976harmonic,srivastava2014mixed,lu2016variational}, are combined. The superior convergence of the mixed method over both PWE and Rayleigh quotient has been theoretically proven in literature \cite{babuska1978numerical} and also been demonstrated in a numerical study \cite{lu2016variational}. Previous implementation of the mixed variation formulation employed numerical integration to evaluate each matrix element in the eigenvalue problem. In this paper it is shown that by expressing the material properties and test functions using Fourier expansion, the need for explicit numerical integration can be mitigated and closed form expressions for the eigenvalue matrices can be achieved unlike the one presented by \citeN{lu2016variational}. It is shown in this paper how employing the Fourier expansion of material properties converts the variational integrals to simple sums. These sums directly represent the matrices of the variational method and the matrix elements are merely the material property Fourier coefficients of appropriate orders. The related matrices are required to be evaluated a limited number of times through the calculation of the discrete Fourier transform (DFT), thus accelerating matrix assembly. The number of Fourier coefficients needed in the matrix assembly is ultimately determined by the number of Fourier expansion terms, therefore, relatively high accuracy results and consistent convergence performance can be achieved by using low material sampling resolution. We describe clearly how the method can be applied to 1-, 2-, and 3-D unit cells of arbitrary complexity in their Bravais structure and in the shape, size, number, and anisotropicity of their micro-constituents. We present 1-, 2- and 3-D test cases which verify the results of the formulation with published results in literature (exact solution for 1-D, plane wave approximation for 2-D and FDTD solution for 3-D). We also investigate the effect of Fourier coefficient resolution on the accuracy of bandstructure calculations and present coparative studies on the convergence and stability behavior of the present method vis-a-vis numerical integration based methods.

\section{Statement of the problem}

\begin{figure}[htp]
\centering
\includegraphics[scale=1]{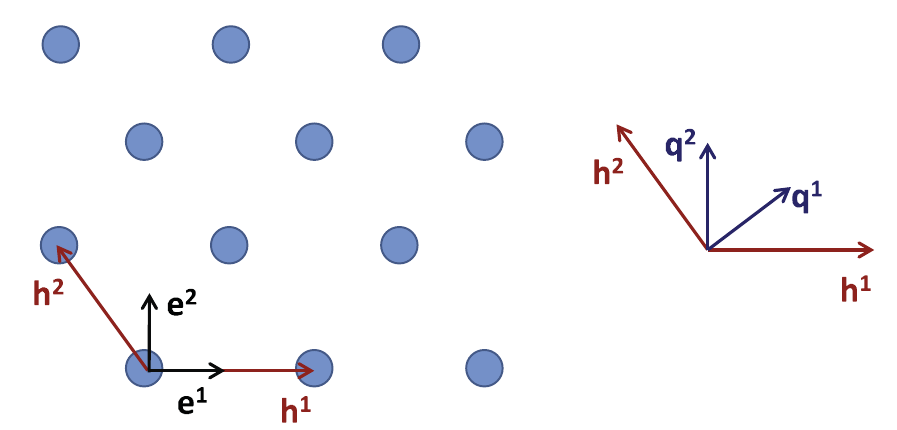}
\caption{Schematic of a 2-D periodic composite. The unit cell vectors ($\mathbf{h}^1,\mathbf{h}^2$), reciprocal basis vectors ($\mathbf{q}^1,\mathbf{q}^2$), and the orthogonal vectors ($\mathbf{e}^1,\mathbf{e}^2$) are shown.}\label{fVectors}
\end{figure}

Consider the problem of elastic wave propagation in a general 3-dimensional periodic composite. The unit cell of the periodic composite is denoted by $\Omega$ and is characterized by 3 base vectors $\mathbf{h}^i$, $i=1,2,3$. Any point within the unit cell can be uniquely specified by the vector $\mathbf{x}=H_i\mathbf{h}^i$ where $0\leq H_i\leq 1,\forall i$. The same point can also be specified in the orthogonal basis as $\mathbf{x}=x_i\mathbf{e}^i$. The reciprocal base vectors of the unit cell are given by:
\begin{equation}
\mathbf{q}^1=2\pi\frac{\mathbf{h}^2\times\mathbf{h}^3}{\mathbf{h}^1\cdot(\mathbf{h}^2\times\mathbf{h}^3)};\quad \mathbf{q}^2=2\pi\frac{\mathbf{h}^3\times\mathbf{h}^1}{\mathbf{h}^2\cdot(\mathbf{h}^3\times\mathbf{h}^1)};\quad \mathbf{q}^3=2\pi\frac{\mathbf{h}^1\times\mathbf{h}^2}{\mathbf{h}^3\cdot(\mathbf{h}^1\times\mathbf{h}^2)},
\end{equation}
such that $\mathbf{q}^i\cdot\mathbf{h}^j=2\pi\delta_{ij}$. Reciprocal lattice vectors are now represented as a linear combination of the reciprocal base vectors, $\mathbf{G}^\mathbf{n}=n_i\mathbf{q}^i$, where $n_i$ are integers. It must be noted that the denominators of the above vectors are merely the volume of the unit cell. Fig. \ref{fVectors} shows the schematic of a 2-D unit cell, clearly indicating the unit cell basis vectors, the reciprocal basis vectors and the orthogonal basis vectors.
The wave-vector for a Bloch-wave traveling in the composite are given as $\mathbf{k}=Q_i\mathbf{q}^i$ where $0\leq Q_i\leq 1,\forall i$. The composite is characterized by a spatially varying stiffness tensor, $C_{jkmn}(\mathbf{x})$, and density, $\rho(\mathbf{x})$, which satisfy the following periodicity conditions:
\begin{equation}
C_{jkmn}(\mathbf{x}+n_i\mathbf{h}^i)=C_{jkmn}(\mathbf{x});\quad \rho(\mathbf{x}+n_i\mathbf{h}^i)=\rho(\mathbf{x}),
\end{equation}
where $n_i(i=1,2,3)$ are integers.

\subsection{Field equations and boundary conditions}

For harmonic elastodynamic problems the equations of motion and kinematic relations at any point $\mathbf{x}$ in $\Omega$ are given by
\begin{equation}\label{equationofmotion}
\sigma_{jk,k}=-\lambda\rho u_j; \quad \varepsilon_{jk}=.5(u_{j,k}+u_{k,j}),
\end{equation}
where $\lambda=\omega^2$, and $\boldsymbol{\sigma}e^{-i\omega t},\boldsymbol{\varepsilon}e^{-i\omega t},\mathbf{u}e^{-i\omega t}$ are the space and time dependent stress tensor, strain tensor, and displacement vector, respectively. The stress tensor is related to the strain tensor through the elasticity tensor, $\sigma_{jk}=C_{jkmn}\varepsilon_{mn}$. The traction and displacement at any point in the composite are related to the corresponding traction and displacement at another point, separated from the first by a unit cell, through Bloch relations. These relations serve as the homogeneous boundary conditions on $\partial\Omega$. If the Bloch wave-vector is $\mathbf{k}$ then these boundary conditions are given by:
\begin{equation}\label{boundaryconditions}
u_j(\mathbf{x}+\mathbf{h}^i)=u_j(\mathbf{x})e^{i\mathbf{k}\cdot\mathbf{h}^i};\quad t_j(\mathbf{x}+\mathbf{h}^i)=-t_j(\mathbf{x})e^{i\mathbf{k}\cdot\mathbf{h}^i}; \quad \mathbf{x}\in\partial\Omega,
\end{equation}
where $t_j=\sigma_{jk}\nu_k$ are the components of the traction vector and $\boldsymbol{\nu}$ is the exterior normal vector on $\partial\Omega$. Under the Bloch boundary conditions the nature of elastic waves in periodic composites is expressed as an infinite set of eigenvalues which depend upon both frequency and wavenumber. This is the phononic dispersion relation of the composite and many numerical techniques have been devised to compute it. Here we focus on combining the desirable properties of the PWE method and the mixed variational method. The former is the method which is easiest to implement and the latter is the one which displays one of the highest convergence rates.

\subsection{Plane Wave Expansion Solution}

The Plane Wave method uses the periodicity of the unit cell to expand the material properties in a Fourier series involving the reciprocal lattice vectors:
\begin{eqnarray}\label{PWEexp}
\displaystyle \rho(\mathbf{x})=\sum_{\mathbf{n}}\rho^\mathbf{n}e^{i\mathbf{G}^\mathbf{n}\cdot\mathbf{x}};\quad \mathbf{C}(\mathbf{x})=\sum_{\mathbf{n}}\mathbf{C}^\mathbf{n}e^{i\mathbf{G}^\mathbf{n}\cdot\mathbf{x}},
\end{eqnarray}
where
\begin{eqnarray}\label{PWEFourier}
\displaystyle \rho^\mathbf{n}=\frac{1}{\Omega}\int_\Omega\rho(\mathbf{x})e^{-i\mathbf{G^n}\cdot\mathbf{x}}d\Omega;\quad \mathbf{C}^\mathbf{n}=\frac{1}{\Omega}\int_\Omega\mathbf{C}(\mathbf{x})e^{-i\mathbf{G^n}\cdot\mathbf{x}}d\Omega.
\end{eqnarray}
Furthermore, it expands the displacement field in a Bloch modified Fourier series involving the reciprocal lattice vectors:
\begin{eqnarray}\label{PWEDisp}
\displaystyle \mathbf{u}(\mathbf{x},t)=e^{i(\mathbf{k}\cdot\mathbf{x}-\omega t)}\sum_\mathbf{n}\mathbf{u}^\mathbf{n}e^{i\mathbf{G^n}\cdot\mathbf{x}}.
\end{eqnarray}
Clearly the above equation automatically satisfies the Bloch periodicity conditions on the displacement field (Eq. \ref{boundaryconditions}). Subsituting the above expansions into the equation of motion (Eq. \ref{equationofmotion}) directly results in an eigenvalue problem whose solutions are the $\mathbf{k},\omega$ pairs which constitute the phononic bandstructure of the composite. The eigenvectors provide the corresponding displacement modeshapes.

\subsection{Mixed variational formulation}

An alternative solution based on the Hu-Washizu variational principle can be formulated for the phononic problem \cite{srivastava2014mixed}. It has been shown by \citeN{minagawa1976harmonic} that the solution to Eq. (\ref{equationofmotion}) that satisfies the boundary conditions, Eq. (\ref{boundaryconditions}), renders the following functional stationary:
\begin{equation}\label{mixedvariational}
\lambda_N=\frac{\langle\sigma_{jk},u_{j,k}\rangle+\langle u_{j,k},\sigma_{jk}\rangle-\langle D_{jkmn}\sigma_{jk},\sigma_{mn}\rangle}{\langle\rho u_j,u_j\rangle},
\end{equation}
where $\mathbf{D}$ is the compliance tensor and the inner product is given by:
\begin{equation}
\langle u,v\rangle=\int_\Omega uv^*d\Omega,
\end{equation}
where $v^*$ is the complex conjugate of $v$. Now the stress and displacement fields are approximated using the following test functions:
\begin{equation}\label{approximation}
\bar{u}_j=\sum_{\mathbf{n}}U^\mathbf{n}_jf^\mathbf{n}(\mathbf{x});\quad \bar{\sigma}_{jk}=\sum_\mathbf{n}S^\mathbf{n}_{jk}f^\mathbf{n}(\mathbf{x}),
\end{equation}
where the test functions are orthogonal in the sense that $\langle f^\mathbf{n},f^{\mathbf{n}'}\rangle$ is proportional to $\delta_{\mathbf{n}\mathbf{n}'}$, $\boldsymbol{\delta}$ being the Kronecker delta. Substituting from Eq. (\ref{approximation}) to Eq. (\ref{mixedvariational}) and setting the derivative of $\lambda_N$ with respect to the unknown coefficients, ($U^\mathbf{n}_j,S^\mathbf{n}_{jk}$), equal to zero, the following system of
linear homogeneous equations can be obtained:
\begin{eqnarray}\label{equationshomogeneous}
\langle\bar{\sigma}_{jk,k}+\lambda_N\rho\bar{u}_j,f^{\mathbf{n}'}\rangle=0\nonumber,\\
\langle D_{jkmn}\bar{\sigma}_{mn}-\bar{u}_{(j,k)},f^{\mathbf{n}'}\rangle=0\nonumber,\\
j,k,m,n=1,2,3 ,
\end{eqnarray}
where $\bar{u}_{(j,k)}\equiv\bar{\varepsilon}_{jk}=.5(\bar{u}_{j,k}+\bar{u}_{k,j})$. While real basis functions can be used in the expansion in Eq. (\ref{approximation}) \cite{lu2016variational}, in this paper trigonometric basis is employed to draw an equivalence with the PWE method. To approximate the stress and displacement fields in Eq. (\ref{approximation}), test functions of the following form are used:
\begin{equation}
f^\mathbf{n}(\mathbf{x})=e^{i(\mathbf{k}\cdot\mathbf{x}+\mathbf{G^n}\cdot\mathbf{x})}.
\end{equation}
The test functions and the field variables clearly satisfy the Bloch boundary conditions. Their orthogonality can be noted from the following:
\begin{equation}\label{transfer}
\langle f^\mathbf{n},f^{\mathbf{n}'}\rangle=\int_\Omega e^{i[\mathbf{G^n}-\mathbf{G^{n'}}]\cdot \mathbf{x}}d\Omega=\Omega\int_0^1\int_0^1\int_0^1 e^{i2\pi[\alpha H_1+\beta H_2+\gamma H_3]}dH_1dH_2dH_3.
\end{equation}
The above holds because the volume element $d\Omega$, when written in the $\mathbf{h}$ coordinate system, involves the square root of the determinant of the Gramian matrix of $\mathbf{h}^i$. This is, in turn, equal to the volume spanned by $\mathbf{h}^i$ which is merely the volume $\Omega$ of the unit cell. Clearly the integral above is zero unless the integers $\alpha,\beta,\gamma$ are all zero confirming the orthogonality of the basis functions. We also note, for subsequent expediency, that the spatial derivatives of the test functions can be expressed by:
\begin{equation}\label{spatialder}
f^{\mathbf{n}}_{,i}=i2\pi K^{\mathbf{n}}_{i}f^{\mathbf{n}},
\end{equation}
where $K^\mathbf{n}_i=T_{ij}(Q_j+n_j)$ and $\mathbf{T}$ is the matrix which takes the vector $\mathbf{x}$ to $\mathbf{H}$:
\begin{equation}
\{H_1 H_2 H_3\}^\mathrm{T}\equiv [\mathbf{T}]\{x_1 x_2 x_3\}^\mathrm{T},
\end{equation}
where $\{ \}^\mathrm{T}$ denotes the transpose operation. The bandstructure of the composite is given by the $\mathbf{k}-\omega$ pairs which lead to nontrivial solutions of Eq. (\ref{equationshomogeneous}). To calculate these pairs Eq. (\ref{equationshomogeneous}) is first written in the following equivalent matrix form:
\begin{eqnarray}\label{equationshomogeneousMatrix}
\nonumber \mathbf{HS}+\lambda_N\mathbf{\Omega U}=0,\\
\mathbf{\Phi S}+\mathbf{H^*U}=0.
\end{eqnarray}
Column vectors $\mathbf{S},\mathbf{U}$ contain the unknown coefficients of the periodic expansions of stress and displacement, respectively. Matrices $\mathbf{H},\mathbf{\Omega},\mathbf{\Phi},\mathbf{H}^*$ contain the integrals of the various functions appearing in Eq. (\ref{equationshomogeneous}). Their sizes depend upon whether the problem under consideration is 1-, 2-, or 3-D. These matrices would be described more clearly in the subsequent sections in which numerical examples are shown. The above system of equations can be recast into the following general form of eigenvalue problem:
\begin{equation}\label{eigenvalueproblem}
\mathbf{H}\mathbf{\Phi}^{-1}\mathbf{H}^*\mathbf{U}=\lambda_N\mathbf{\Omega}\mathbf{U},
\end{equation}
whose eigenvalue solutions represent the frequencies ($\omega_N=\sqrt{\lambda_N}$) associated with the wave-vector under consideration ($\mathbf{k}$). The eigenvectors of the above equation are used to calculate the displacement modeshapes from Eq. (\ref{approximation}). The relation $\mathbf{S}=-\mathbf{\Phi}^{-1}\mathbf{H}^*\mathbf{U}$ is used to evaluate the stress eigenvector which is subsequently used to calculate the stress modeshape from Eq. (\ref{approximation}).

\subsection{Employing Fourier Expansion of Material Properties}
By combining Eq. (\ref{equationshomogeneous}) with the orthogonality condition on the test function and the spatial derivative relation, Eq. (\ref{spatialder}), closed form expressions for the following integrals can be written as:
\begin{equation}
\langle \bar{\sigma}_{jk,k},f^{\mathbf{n}'}\rangle=i2\pi \Omega K^{\mathbf{n}}_kS^{\mathbf{n}}_{jk};\quad \langle \bar{u}_{(j,k)},f^{\mathbf{n}'}\rangle=i\pi\Omega [K^{\mathbf{n}}_kU^{\mathbf{n}}_{j}+K^{\mathbf{n}}_jU^{\mathbf{n}}_{k}].
\end{equation}
The other two sets of integrals in Eq. (\ref{equationshomogeneous}) involve the material properties of the unit cell and cannot be immediately written down in closed form as above. However, closed form expressions can be generated by first expanding the material properties themselves in a Fourier series:
\begin{eqnarray}\label{MVexp}
\displaystyle \rho(\mathbf{x})=\sum_{\mathbf{m}}\rho^\mathbf{m}e^{i\mathbf{G}^\mathbf{m}\cdot\mathbf{x}};\quad \mathbf{C}(\mathbf{x})=\sum_{\mathbf{m}}\mathbf{C}^\mathbf{m}e^{i\mathbf{G}^\mathbf{m}\cdot\mathbf{x}}.
\end{eqnarray}
Now the integrals can be simplified:
\begin{eqnarray}
\nonumber\displaystyle \langle \lambda_N\rho\bar{u}_j,f^{\mathbf{n}'}\rangle=\lambda_N\int_\Omega \left[\sum_{\mathbf{m}}\rho^\mathbf{m}e^{i\mathbf{G}^\mathbf{m}\cdot\mathbf{x}}\right]\left[\sum_{\mathbf{n}}U_j^\mathbf{n}e^{i\mathbf{G}^\mathbf{n}\cdot\mathbf{x}}\right]e^{-i\mathbf{G}^{\mathbf{n}'}\cdot\mathbf{x}}d\Omega\\
=\lambda_N\Omega\sum_\mathbf{n}\rho^{\mathbf{n}-\mathbf{n}'}U^\mathbf{n}_j.
\end{eqnarray}
Similarly:
\begin{eqnarray}
\displaystyle \langle D_{jkmn}\bar{\sigma}_{mn},f^{\mathbf{n}'}\rangle=\Omega\sum_\mathbf{n}D_{jkmn}^{\mathbf{n}-\mathbf{n}'}S_{mn}^\mathbf{n}.
\end{eqnarray}
The above equations are the closed form expressions for all the integrals required to formulate the phononic eigenvalue problem under the mixed variational method. These expressions require no numerical integration and can be easily computed by employing the fast Fourier transform algorithms readily available with commercial software. We also note that these expressions make clear that the resolution with which the unit cell needs to be discretized for sampling the material properties is ultimately governed by the number of Fourier terms used in the expansion of the field variables. This is evident from the independence of the final equations from $\mathbf{m}$. However, the accuracy of the Fourier coefficients which contribute to the matrix elements depends on the sampling resolution of material properties. Increasing the sampling resolution of the material properties beyond a certain limit, as governed by the highest order Fourier term necessary, has minimal effect on improving the accuracy of the solution. In the subsequent sections we adapt the above expression for 1-, 2-, and 3-D and give explicit formulae for the matrices involved in the mixed method.

\section{1-D phononic composites}

There is only one possible Bravais lattice in 1-D with a unit cell vector whose length equals the length of the unit cell itself. Without any loss of generality, we take the direction of this vector to be the same as $\mathbf{e}^1$. If the length of the unit cell is $a$, then we have $\mathbf{h}^1=a\mathbf{e}^1$ and $\mathbf{x}=H_1\mathbf{h}^1$. The reciprocal vector is given by $\mathbf{q}^1=(2\pi/a)\mathbf{e}^1$. The wave-vector of a Bloch wave traveling in this composite is specified as $\mathbf{k}=Q_1\mathbf{q}^1$. To completely characterize the bandstructure of the unit cell it is sufficient to evaluate the dispersion relation in the irreducible Brillouin zone ($-.5\leq Q_1\leq .5$). For plane longitudinal waves propagating in the $\mathbf{e}^1$ direction, the relevant material properties are the compliance $D(\mathbf{x})$ and density $\rho(\mathbf{x})$ which are both periodic with the unit cell. In this case the only displacement component of interest is $u_1$ and the only relevant stress component is $\sigma_{11}$ (for plane shear waves traveling in $\mathbf{e}^1$ direction the quantities of interest are $u_2$ and $\sigma_{12}$). The Fourier term exponents forming the test function in 1-D are given by $\mathbf{G}^\mathbf{n}\equiv\mathbf{G}^{n_1}=n_1\mathbf{q}^1$ where $n_1$ is an integer which varies from some $-M$ to $M$. The displacement and stress are now expressed as:
\begin{equation}\label{approximation1d}
\bar{u}_1=\sum_{n_1=-M}^MU^{n_1}_1e^{i\left[\mathbf{k}+\mathbf{G}^{n_1}\right]\cdot\mathbf{x}};\quad \bar{\sigma}_{11}=\sum_{n_1=-M}^MS^{n_1}_{11}e^{i\left[\mathbf{k}+\mathbf{G}^{n_1}\right]\cdot\mathbf{x}}.
\end{equation}
The Fourier coefficients of density and compliance are given by:
\begin{eqnarray}\label{fouriercoef1d}
\nonumber \rho^m=\frac{1}{\Omega}\int_\Omega\rho(\mathbf{x})e^{-i\mathbf{G}^m\cdot\mathbf{x}}d\mathbf{x}=\int_0^1\rho(aH_1)e^{-i2\pi mH_1}dH_1,\\
D^m=\frac{1}{\Omega}\int_\Omega D(\mathbf{x})e^{-i\mathbf{G}^m\cdot\mathbf{x}}d\mathbf{x}=\int_0^1D(aH_1)e^{-i2\pi mH_1}dH_1.
\end{eqnarray}
The eigenvalue problem Eq. (\ref{equationshomogeneousMatrix}) involves the following column vectors:
\begin{eqnarray}
\mathbf{U}=\{U^{-M}_1\;...\;U^{0}_1\;...\;U^{M}_1\}^T\nonumber,\\
\mathbf{S}=\{S^{-M}_{11}\;...\;S^{0}_{11}\;...\;S^{M}_{11}\}^T,
\end{eqnarray}
and the associated coefficient matrices have the following nonzero elements (0 indexing assumed):
\begin{eqnarray}\label{coefficients1d}
\nonumber[\mathbf{H}]_{I_1I_1}=i2\pi(Q_1+n_1);\quad [\mathbf{H}]^*_{I_1I_1}=-i2\pi(Q_1+n_1),\\
\nonumber[\mathbf{\Omega}]_{I_1J_1}=a\rho^{n_1-n'_1};\quad \nonumber[\mathbf{\Phi}]_{I_1J_1}=aD^{n_1-n'_1},\\
\nonumber I_1=n_1+M,\;J_1=n'_1+M,\\
n_1,n'_1=-M,...,M.
\end{eqnarray}
Now the phononic eigenvalue problem can be solved for the frequencies $\omega_N$ which correspond to an assumed value of $Q_1$ through Eq. (\ref{eigenvalueproblem}).

\section{2-D phononic composites}

There are five possible Bravais lattices in 2-D. However, they can be specified using two unit cell vectors ($\mathbf{h}^1,\mathbf{h}^2$). The reciprocal vectors are $\mathbf{q}^1,\mathbf{q}^2$ and the area of the unit cell is $A$. The wave-vector of a Bloch wave traveling in this composite is specified as $\mathbf{k}=Q_1\mathbf{q}^1+Q_2\mathbf{q}^2$. To characterize the bandstructure of the unit cell we evaluate the dispersion relation along the boundaries of the irreducible Brillouin zone ($0\leq Q_1\leq .5,Q_2=0;\;Q_1=.5,0\leq Q_2\leq .5;\;0\leq Q_1\leq .5,Q_2=Q_1$). In traditional notation these boundaries are specified as $\Gamma-X,X-M,M-\Gamma$, respectively. The relevant stress components for the plane strain case are $\sigma_{11},\sigma_{22},\sigma_{12}$ and the relevant displacement components are $u_1,u_2$. For an isotropic material in plane strain the compliance tensor $\mathbf{D}$ is given by:
\begin{equation}
D_{jkmn}=\frac{1}{2\mu}\left[\frac{1}{2}(\delta_{jm}\delta_{kn}+\delta_{jn}\delta_{km})-\frac{\lambda}{2(\mu+\lambda)}\delta_{jk}\delta_{mn}\right];\quad j,k,m,n=1,2,
\end{equation}
where $\lambda,\mu$ are the Lam\'{e} constants of the material. The Fourier term exponents forming the test function in 2-D are given by $\mathbf{G}^\mathbf{n}=n_1\mathbf{q}^1+n_2\mathbf{q}^2$ where $n_1,n_2$ are integers which vary from some $-M$ to $M$. The stresses and displacements are approximated by the following 2-D periodic functions:
\begin{equation}\label{approximation2d}
\bar{u}_i=\sum_{n_1,n_2=-M}^MU^{\mathbf{n}}_ie^{i\left[\mathbf{k}+\mathbf{G}^{\mathbf{n}}\right]\cdot\mathbf{x}};\quad \bar{\sigma}_{ij}=\sum_{n_1,n_2=-M}^MS^{\mathbf{n}}_{ij}e^{i\left[\mathbf{k}+\mathbf{G}^{\mathbf{n}}\right]\cdot\mathbf{x}};\quad i,j=1,2,
\end{equation}
where the superscript $\mathbf{n}$ refers to the ordered pair ($n_1,n_2$). The Fourier coefficients of density and compliance are given by:
\begin{eqnarray}\label{fouriercoef2D}
\nonumber \rho^\mathbf{m}=\int_0^1\int_0^1\rho(H_i\mathbf{h}^i)e^{-i2\pi [m_1H_1+m_2H_2]}dH_1dH_2,\\
\mathbf{D}^\mathbf{m}=\int_0^1\int_0^1\mathbf{D}(H_i\mathbf{h}^i)e^{-i2\pi [m_1H_1+m_2H_2]}dH_1dH_2.
\end{eqnarray}
The matrix form of the eigenvalue problem is given by Eq. (\ref{equationshomogeneousMatrix}) with the following column vectors:
\begin{equation}
\mathbf{U}=\{U^{\mathbf{n}}_1\;U^{\mathbf{n}}_2\}^T\nonumber;\quad \mathbf{S}=\{S^{\mathbf{n}}_{11}\;S^{\mathbf{n}}_{22}\;S^{\mathbf{n}}_{12}\}^T.
\end{equation}
The length of the column vector $\mathbf{U}$ is $2(2M+1)^2$ and the length of $\mathbf{S}$ is $3(2M+1)^2$. Corresponding to these column vectors, the size of $\mathbf{H}$ is $3(2M+1)^2\times 2(2M+1)^2$, $\mathbf{\Omega}$ is $2(2M+1)^2\times 2(2M+1)^2$, and $\mathbf{\Phi}$ is $3(2M+1)^2\times 3(2M+1)^2$. To clarify the contents of the matrices $[\mathbf{H}],[\mathbf{\Omega}],[\mathbf{\Phi}]$ we introduce the following modified coordinates ($n'_1,n'_2=-M...,M$):
\begin{eqnarray}
\nonumber I_1=(n_1+M)(2M+1)+(n_2+1+M);\quad
J_1=(n'_1+M)(2M+1)+(n'_2+1+M)\\
\nonumber I_2=I_1+(2M+1)^2;\;J_2=J_1+(2M+1)^2\\
\nonumber I_3=I_2+(2M+1)^2;\;J_3=J_2+(2M+1)^2.
\end{eqnarray}
Components of the $\mathbf{H}$ matrix are given by:
\begin{equation}\label{eh}
[\mathbf{H}]_{I_1J_1}=i2\pi AK^{\mathbf{n}}_1;\quad [\mathbf{H}]_{I_2J_2}=i2\pi AK^{\mathbf{n}}_2;\quad
[\mathbf{H}]_{I_1J_3}=[\mathbf{H}]_{I_2J_2};\quad [\mathbf{H}]_{I_2J_3}=[\mathbf{H}]_{I_1J_1}.
\end{equation}
We also have $[\mathbf{H}]^*=-[\mathbf{H}]^T$ where the superscript $T$ denotes a matrix transpose. Components of the $\mathbf{\Omega}$ matrix are given by:
\begin{equation}\label{eo}
[\mathbf{\Omega}]_{I_1J_1}=A\rho^{\mathbf{n}-\mathbf{n}'};\quad[\mathbf{\Omega}]_{I_2J_2}=[\mathbf{\Omega}]_{I_1J_1},
\end{equation}
with the rest of the terms in the $\mathbf{\Omega}$ matrix being zero. The components of the $\mathbf{\Phi}$ matrix are given by:
\begin{eqnarray}\label{ep}
\nonumber [\mathbf{\Phi}]_{I_1J_1}=AD_{1111}^{\mathbf{n}-\mathbf{n}'}\quad [\mathbf{\Phi}]_{I_1J_2}=AD_{1122}^{\mathbf{n}-\mathbf{n}'}\quad [\mathbf{\Phi}]_{I_1J_3}=2AD_{1112}^{\mathbf{n}-\mathbf{n}'}\\
\nonumber [\mathbf{\Phi}]_{I_2J_1}=AD_{2211}^{\mathbf{n}-\mathbf{n}'}\quad [\mathbf{\Phi}]_{I_2J_2}=AD_{2222}^{\mathbf{n}-\mathbf{n}'}\quad [\mathbf{\Phi}]_{I_2J_3}=2AD_{2212}^{\mathbf{n}-\mathbf{n}'}\\
\left[\mathbf{\Phi}\right]_{I_3J_1}=2AD_{1211}^{\mathbf{n}-\mathbf{n}'}\quad [\mathbf{\Phi}]_{I_3J_2}=2AD_{1222}^{\mathbf{n}-\mathbf{n}'}\quad [\mathbf{\Phi}]_{I_3J_3}=4AD_{1212}^{\mathbf{n}-\mathbf{n}'}.
\end{eqnarray}

\section{3-D phononic composites}

Similar expressions can be derived for the 3-D case. There are three unit cell vectors in 3-D ($\mathbf{h}^1,\mathbf{h}^2,\mathbf{h}^3$) and three reciprocal vectors ($\mathbf{q}^1,\mathbf{q}^2,\mathbf{q}^3$). Volume of the unit cell is $V$. The wave-vector of a Bloch wave traveling in this composite is specified as $\mathbf{k}=Q_i\mathbf{q}^i,i=1,2,3$. The Fourier term exponents forming the test function in 3-D are given by $\mathbf{G}^\mathbf{n}=n_i\mathbf{q}^i$ where $n_1,n_2,n_3$ are integers which vary from some $-M$ to $M$. The stresses and displacements are approximated by the following 3-D periodic functions:
\begin{equation}\label{approximation3d}
\bar{u}_i=\sum_{n_1,n_2,n_3=-M}^MU^{\mathbf{n}}_ie^{i\left[\mathbf{k}+\mathbf{G}^{\mathbf{n}}\right]\cdot\mathbf{x}},\quad \bar{\sigma}_{ij}=\sum_{n_1,n_2,n_3=-M}^MS^{\mathbf{n}}_{ij}e^{i\left[\mathbf{k}+\mathbf{G}^{\mathbf{n}}\right]\cdot\mathbf{x}};\quad i,j=1,2,3,
\end{equation}
where the superscript $\mathbf{n}$ refers to the ordered pair ($n_1,n_2,n_3$). The Fourier coefficients of density and compliance are given by:
\begin{eqnarray}\label{fouriercoef3D}
\nonumber \displaystyle \rho^\mathbf{m}=\int_0^1\int_0^1\int_0^1\rho(H_i\mathbf{h}^i)e^{-i2\pi [m_1H_1+m_2H_2+m_3H_3]}dH_1dH_2dH_3,\\
\mathbf{D}^\mathbf{m}=\int_0^1\int_0^1\int_0^1\mathbf{D}(H_i\mathbf{h}^i)e^{-i2\pi [m_1H_1+m_2H_2+m_3H_3]}dH_1dH_2dH_3.
\end{eqnarray}
The matrix form of the eigenvalue problem is given by Eq. (\ref{equationshomogeneousMatrix}) with the following column vectors:
\begin{equation}
\mathbf{U}=\{U^{\mathbf{n}}_1\;U^{\mathbf{n}}_2\;U^{\mathbf{n}}_3\}^T; \quad \mathbf{S}=\{S^{\mathbf{n}}_{11}\;S^{\mathbf{n}}_{22}\;S^{\mathbf{n}}_{33}\;S^{\mathbf{n}}_{12}\;S^{\mathbf{n}}_{23}\;S^{\mathbf{n}}_{13}\}^T.
\end{equation}
At this point we introduce the following modified coordinates:
\begin{eqnarray*}
I_1=(n_1+M)M_p^2+(n_2+M)M_p+(n_3+1+M),\\ 
J_1=(n'_1+M)M_p^2+(n'_2+M)M_p+(n'_3+1+M),
\end{eqnarray*}
where $n'_1,n'_2,n'_3=-M...,M$ and $I_i=I_{i-1}+M_p^3$ and $J_i=J_{i-1}+M_p^3$ for $i=2...,6$ and $M_p=2M+1$.
Components of the $\mathbf{H}$ matrix are given by:
\begin{eqnarray}
\nonumber [\mathbf{H}]_{I_1J_1}=i2\pi VK^{\mathbf{n}}_1;\quad [\mathbf{H}]_{I_1J_4}=i2\pi VK^{\mathbf{n}}_2;\quad [\mathbf{H}]_{I_1J_6}=i2\pi VK^{\mathbf{n}}_3\\
\nonumber [\mathbf{H}]_{I_2J_4}=[\mathbf{H}]_{I_3J_6}=[\mathbf{H}]_{I_1J_1}\\
\nonumber [\mathbf{H}]_{I_2J_2}=[\mathbf{H}]_{I_3J_5}=[\mathbf{H}]_{I_1J_4}\\ 
\left[\mathbf{H}\right]_{I_2J_5}=[\mathbf{H}]_{I_3J_3}=[\mathbf{H}]_{I_1J_6}.
\end{eqnarray}
Components of the $\mathbf{\Omega}$ matrix are given by:
\begin{equation}
[\mathbf{\Omega}]_{I_1J_1}=[\mathbf{\Omega}]_{I_2J_2}=[\mathbf{\Omega}]_{I_3J_3}=V\rho^{\mathbf{n}-\mathbf{n}'}.
\end{equation}
The components of the $\mathbf{\Phi}$ matrix are given by:
\begin{eqnarray}
\nonumber[\mathbf{\Phi}]_{I_1J_1}=VD_{1111}^{\mathbf{n}-\mathbf{n}'}\;[\mathbf{\Phi}]_{I_1J_2}=VD_{1122}^{\mathbf{n}-\mathbf{n}'}\;[\mathbf{\Phi}]_{I_1J_3}=VD_{1133}^{\mathbf{n}-\mathbf{n}'}\;[\mathbf{\Phi}]_{I_1J_4}=2VD_{1112}^{\mathbf{n}-\mathbf{n}'}\\
\nonumber [\mathbf{\Phi}]_{I_1J_5}=2VD_{1123}^{\mathbf{n}-\mathbf{n}'}\; [\mathbf{\Phi}]_{I_1J_6}=2VD_{1113}^{\mathbf{n}-\mathbf{n}'}\;[\mathbf{\Phi}]_{I_2J_1}=VD_{2211}^{\mathbf{n}-\mathbf{n}'}\;[\mathbf{\Phi}]_{I_2J_2}=VD_{2222}^{\mathbf{n}-\mathbf{n}'}\\
\nonumber [\mathbf{\Phi}]_{I_2J_3}=VD_{2233}^{\mathbf{n}-\mathbf{n}'}\;[\mathbf{\Phi}]_{I_2J_4}=2VD_{2212}^{\mathbf{n}-\mathbf{n}'}\; [\mathbf{\Phi}]_{I_2J_5}=2VD_{2223}^{\mathbf{n}-\mathbf{n}'}\; [\mathbf{\Phi}]_{I_2J_6}=2VD_{2213}^{\mathbf{n}-\mathbf{n}'}\\
\nonumber[\mathbf{\Phi}]_{I_3J_1}=VD_{3311}^{\mathbf{n}-\mathbf{n}'}\;[\mathbf{\Phi}]_{I_3J_2}=VD_{3322}^{\mathbf{n}-\mathbf{n}'}\;[\mathbf{\Phi}]_{I_3J_3}=VD_{3333}^{\mathbf{n}-\mathbf{n}'}\; [\mathbf{\Phi}]_{I_3J_4}=2VD_{3312}^{\mathbf{n}-\mathbf{n}'}\\
\nonumber [\mathbf{\Phi}]_{I_3J_5}=2VD_{3323}^{\mathbf{n}-\mathbf{n}'}\; [\mathbf{\Phi}]_{I_3J_6}=2VD_{3313}^{\mathbf{n}-\mathbf{n}'}\;[\mathbf{\Phi}]_{I_4J_1}=2VD_{1211}^{\mathbf{n}-\mathbf{n}'}\;[\mathbf{\Phi}]_{I_4J_2}=2VD_{1222}^{\mathbf{n}-\mathbf{n}'}\\
\nonumber [\mathbf{\Phi}]_{I_4J_3}=2VD_{1233}^{\mathbf{n}-\mathbf{n}'}\;[\mathbf{\Phi}]_{I_4J_4}=4VD_{1212}^{\mathbf{n}-\mathbf{n}'}\; [\mathbf{\Phi}]_{I_4J_5}=4VD_{1223}^{\mathbf{n}-\mathbf{n}'}\; [\mathbf{\Phi}]_{I_4J_6}=4VD_{1213}^{\mathbf{n}-\mathbf{n}'}\\
\nonumber[\mathbf{\Phi}]_{I_5J_1}=2VD_{2311}^{\mathbf{n}-\mathbf{n}'}\;[\mathbf{\Phi}]_{I_5J_2}=2VD_{2322}^{\mathbf{n}-\mathbf{n}'}\;[\mathbf{\Phi}]_{I_5J_3}=2VD_{2333}^{\mathbf{n}-\mathbf{n}'}\; [\mathbf{\Phi}]_{I_5J_4}=4VD_{2312}^{\mathbf{n}-\mathbf{n}'}\\
\nonumber [\mathbf{\Phi}]_{I_5J_5}=4VD_{2323}^{\mathbf{n}-\mathbf{n}'}\; [\mathbf{\Phi}]_{I_5J_6}=4VD_{2313}^{\mathbf{n}-\mathbf{n}'}\;[\mathbf{\Phi}]_{I_6J_1}=2VD_{1311}^{\mathbf{n}-\mathbf{n}'}\;[\mathbf{\Phi}]_{I_6J_2}=2VD_{1322}^{\mathbf{n}-\mathbf{n}'}\\
\left[\mathbf{\Phi}\right]_{I_6J_3}=2VD_{1333}^{\mathbf{n}-\mathbf{n}'}\;[\mathbf{\Phi}]_{I_6J_4}=4VD_{1312}^{\mathbf{n}-\mathbf{n}'}\; [\mathbf{\Phi}]_{I_6J_5}=4VD_{1323}^{\mathbf{n}-\mathbf{n}'}\;[\mathbf{\Phi}]_{I_6J_6}=4VD_{1313}^{\mathbf{n}-\mathbf{n}'}.
\end{eqnarray}

\section{Connection with Discrete Fourier Transforms}

In the previous sections we have presented closed form expressions for the matrices involved in the mixed variational method. These equations depend upon the calculation of the Fourier transforms of material properties in 1-, 2-, and 3-D. In this section we show how existing discrete Fourier transform algorithms available for MATLAB and NumPy can be used to easily determine these Fourier coefficients. The correspondence between discrete Fourier transform coefficients and the Fourier coefficients used in the previous sections will be clarified.
\begin{figure}
\centering
\includegraphics[scale=.5]{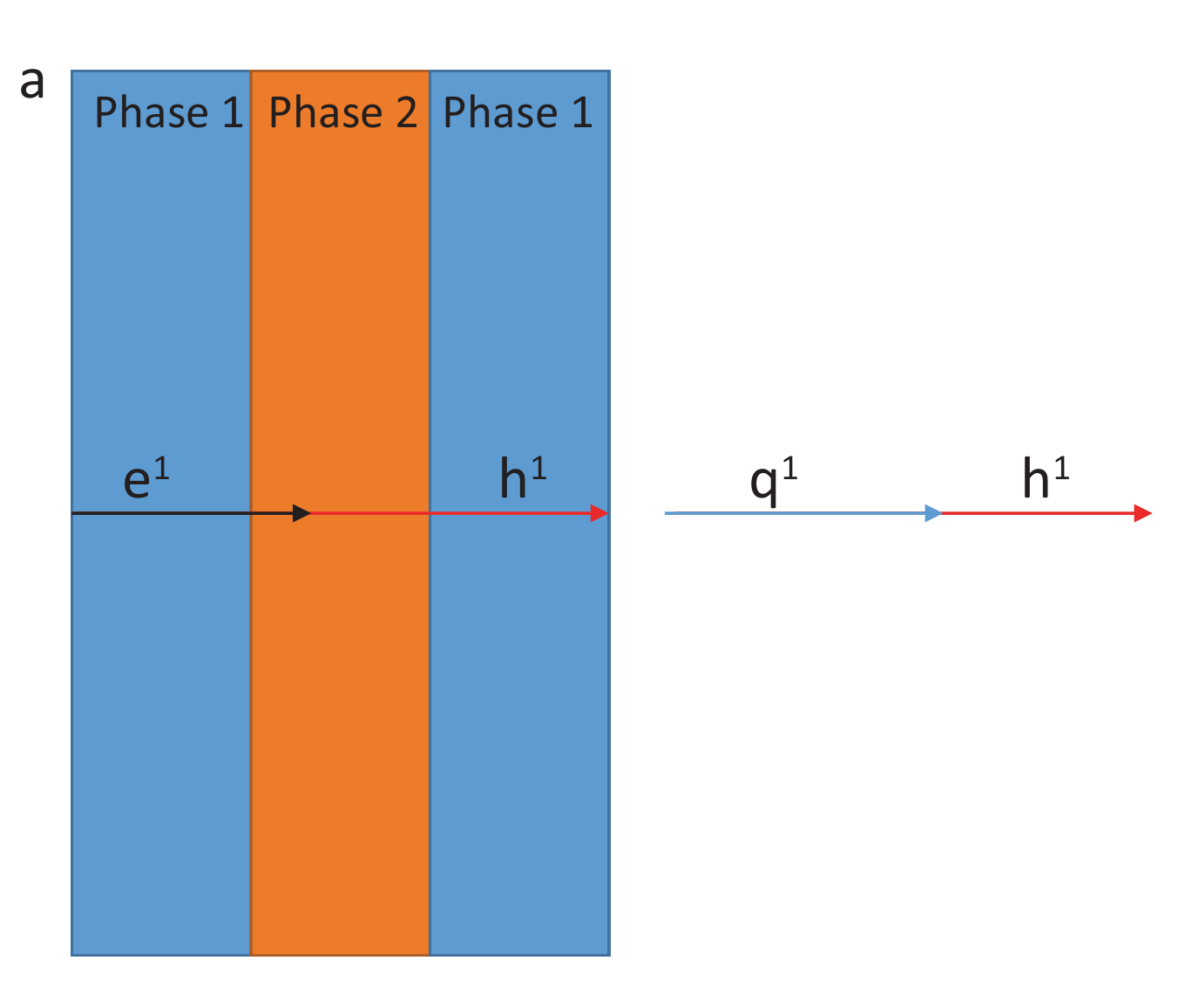}
\includegraphics[scale=.5]{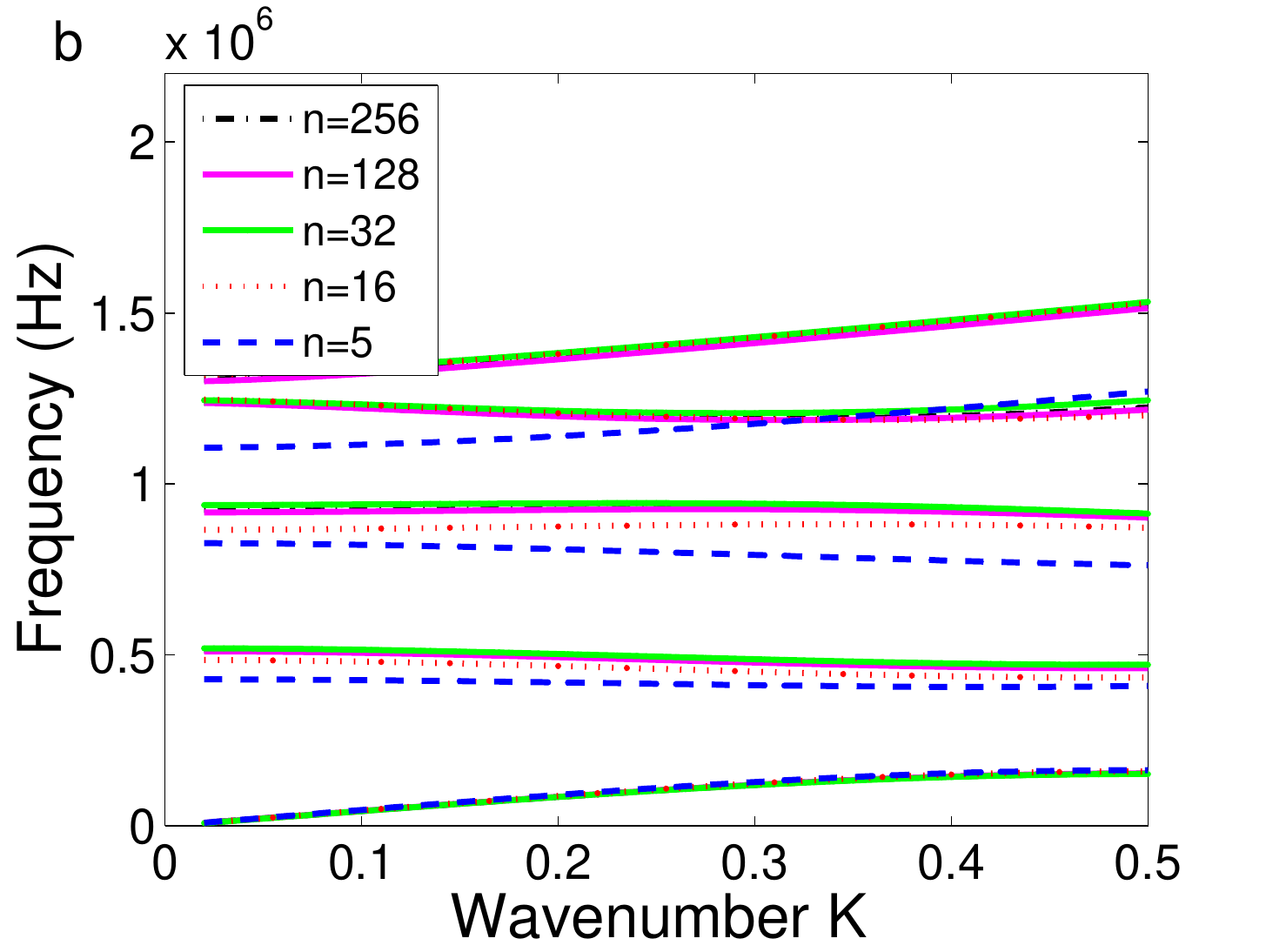}
\caption{a. Schematic of the 1-D layered composite, b. Bandstructure calculation results using the mixed variational formulation with different number of material property sampling points.}\label{fCompLayered}
\end{figure}
\subsection{Layered composites}
In 1-D, the Fourier coefficients of density and compliance are given in Eq. (\ref{fouriercoef1d}). The unit cell is discretized into $p-1$ intervals by sampling it at $p$ points. Now, the $m$th Fourier coefficient of density can be obtained by using the discrete Fourier transform: 
\begin{equation}
\rho^m=\frac{1}{p\Delta x_1}\sum_{u=1}^{p}\rho_ue^{-iG_1^mx_1}\Delta x_1=\frac{1}{p}\sum_{u=1}^{p}\rho_ue^{-im\frac{2\pi}{p}u}=\frac{1}{p}\rho^m_{DFT},
\end{equation}
where $\rho_u$ is the sampled density value, $G_1^m=m\frac{2\pi}{p\Delta x_1}$, and $x_1=u\Delta x_1$. $\rho^m_{DFT}$ is the Fourier coefficient calculated directly from either MATLAB's or NumPy's fft command. The Fourier coefficients of compliance can be similarly written as:
\begin{equation}
D^m=\frac{1}{p\Delta x_1}\sum_{u=1}^{p}D_ue^{-iG_1^mx_1}\Delta x_1=\frac{1}{p}\sum_{u=1}^{p}D_ue^{-im\frac{2\pi}{p}u}=\frac{1}{p}D^m_{DFT}.
\end{equation}

\begin{figure}
\centering
\includegraphics[scale=.5]{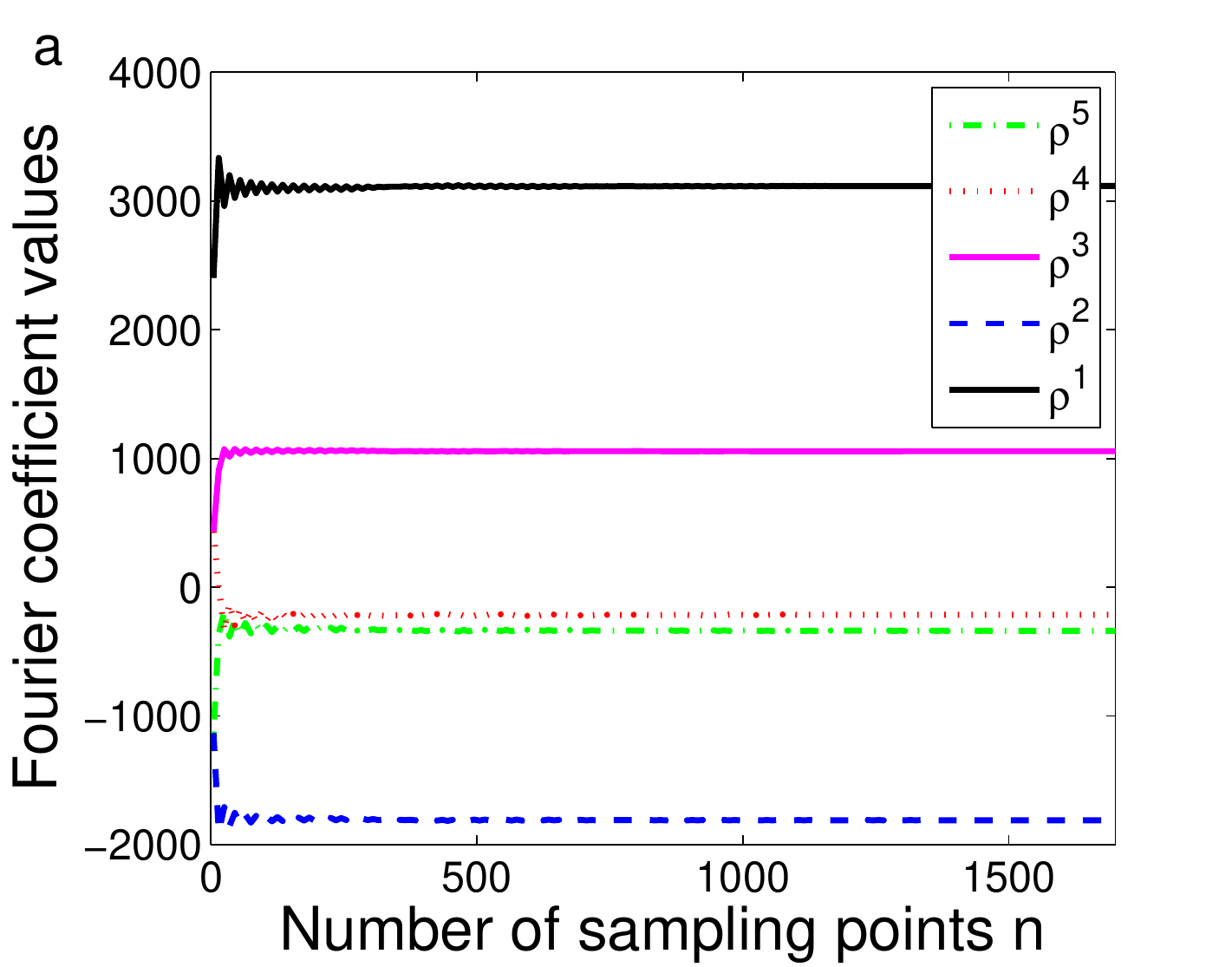}
\includegraphics[scale=.5]{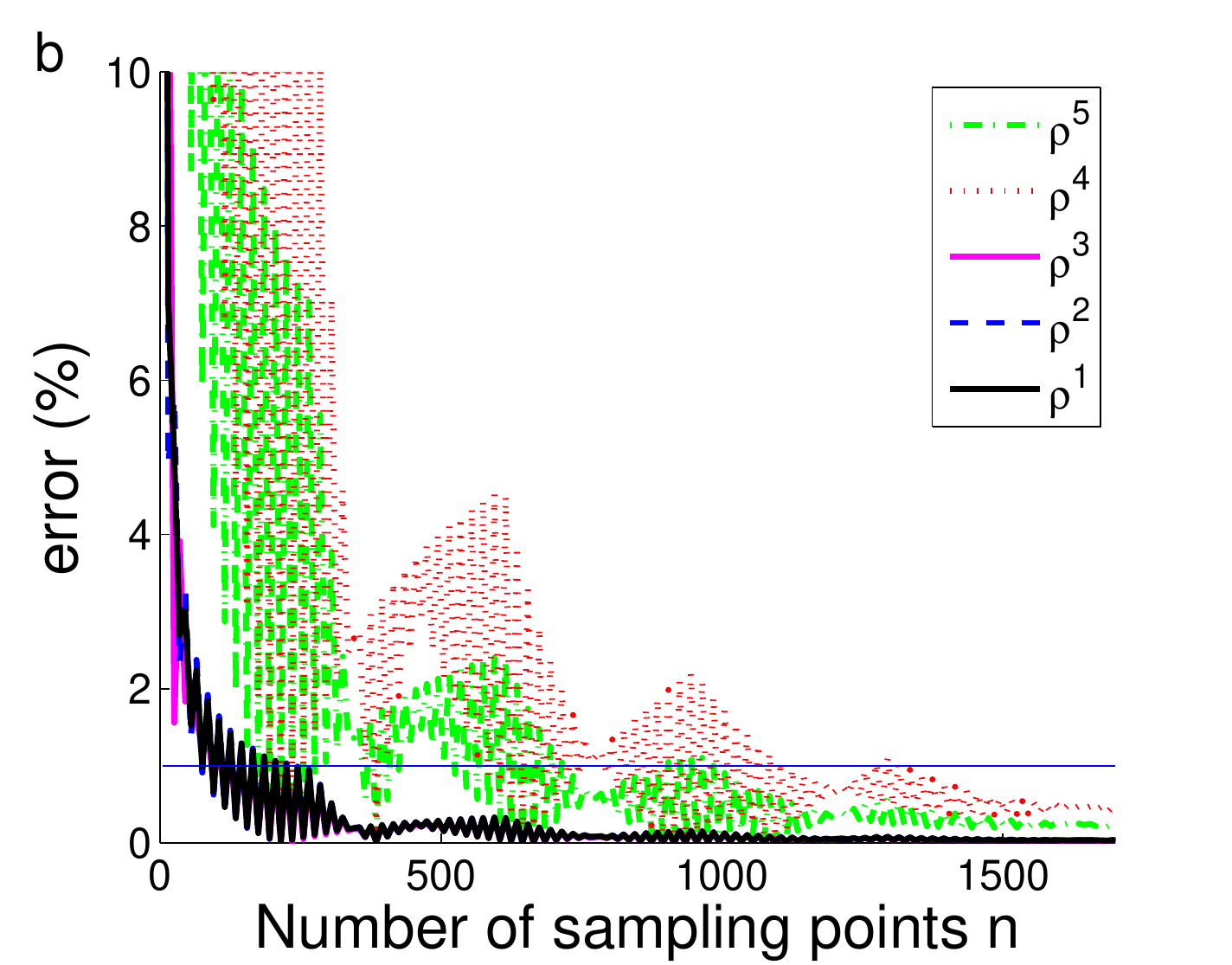}
\caption{a. The values of the first 5 Fourier coefficients using different number of sampling points, b. The percentage errors of the first 5 Fourier coefficients. The first 3 Fourier coefficients converge to less than $1\%$ error around $n=256$ and other coefficients need more than 1000 to converge to $1\%$.}\label{fCompCoeff}
\end{figure}

As an example, we consider a 1-D layered composite with the following material properties:
\begin{enumerate}
\item Phase 1: $E_1=8GPa,\; \rho_1=1000kg/m^3,\; thickness=0.003m$

\item Phase 2: $E_2=300GPa,\; \rho_2=8000kg/m^3,\; thickness=0.0013m$ 
\end{enumerate}
The schematic is shown in Fig. \ref{fCompLayered}a. Here the bandstructure is calculated using $M=2$ which corresponds to five Fourier terms (Eq. \ref{approximation1d}). For $M=2$ only five Fourier coefficients of the material properties contribute to the eigenvalue matrix (Eq. \ref{coefficients1d}). Although only five Fourier coefficients contribute to the solution, their accurate evaluation still depends upon the material property sampling resolution. As shown in Fig. \ref{fCompCoeff}b, the lower Fourier coefficients ($\rho^1$ to $\rho^3$) only require 256 sampling points to converge to less than $1\%$ error. However, more sampling points are needed to achieve similar accuracy for higher coefficients ($\rho^4$ and $\rho^5$). Once enough accuracy is achieved for all the required Fourier coefficients as defined by Eq. (\ref{coefficients1d}), further increasing the sampling resolution has minimal effect on the accuracy of the eigenvalues. 

\begin{figure}
\centering
\includegraphics[scale=.6]{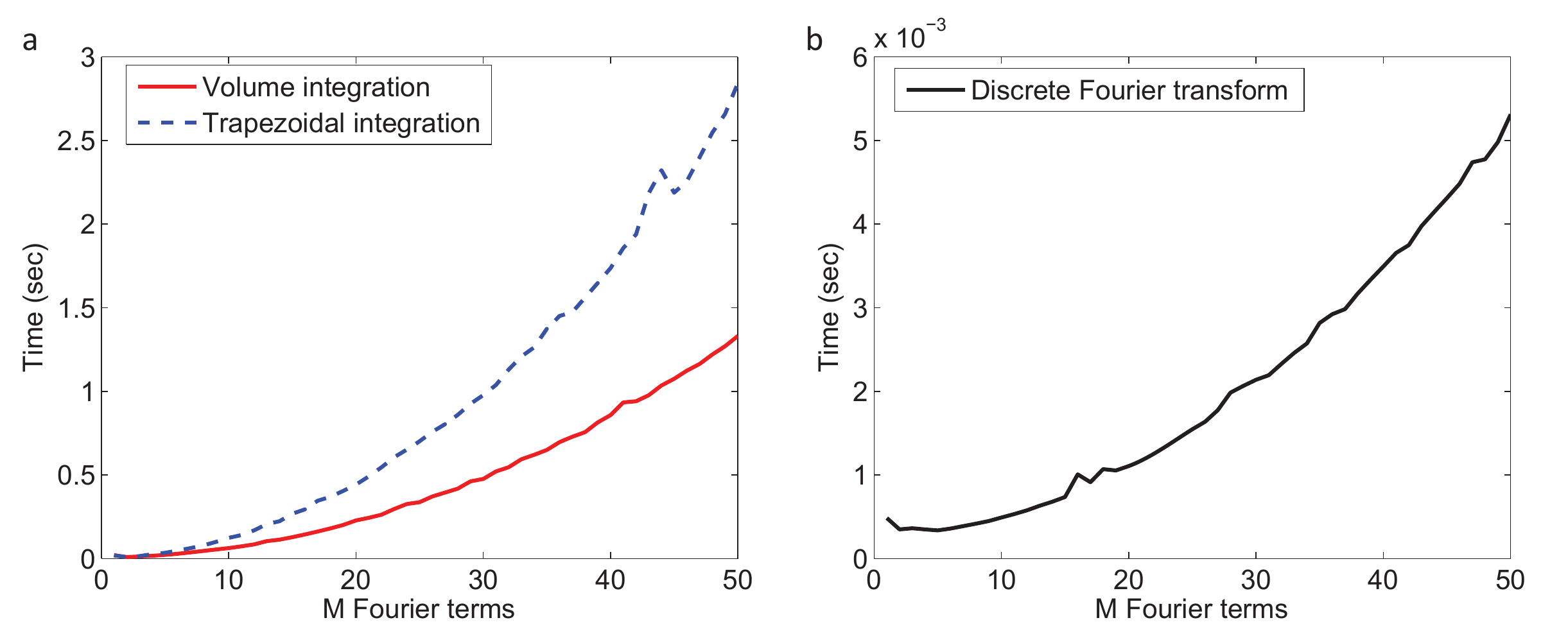}
\caption{Matrix assembling time for a single wave-vector when using: a. numerical integration and b. discrete Fourier transform.}\label{fCompTime}
\end{figure}

The efficiency of calculating phononic bandstructure depends not only on the eigenvalue solver but also on the matrix assembly process. We measure the matrix assembly time, which includes numerical integration, FFT and matrix elements allocation for $\mathbf{H}$, $\mathbf{\Phi}$ and $\mathbf{Omega}$, for a single wave-vector using numerical integration over 256 elements and discrete Fourier transform over the same number of material property sampling points. When the $M$ value is increased, the number of matrix elements grows quadratically for the 1-D case. This results in longer total assembly time, as shown in Fig. \ref{fCompTime}. However, the matrix assembly time is significantly shorter, compared to both centroid based volume integration and higher order trapezoidal integration when discrete Fourier transform is used. The material sampling time using FEniCS is 0.035 second comparing to 0.015 second when mesh and material properties are read for numerical integration. This small advantage does not compensate for the inefficiency during matrix assembly process. 

\subsection{Example: hexagonal unit cell}
In 2-D, the Fourier coefficients of density and compliance, given in Eq. (\ref{fouriercoef2D}), are in the form of two dimensional arrays. The material properties are sampled using $p$ and $q$ points along the $\mathbf{h}^1$ and $\mathbf{h}^2$ directions. The $(m,n)$th Fourier coefficient of density can be obtained by using discrete Fourier transform:
\begin{eqnarray}
\nonumber \rho^{m,n}=\frac{1}{p\Delta x_1 q\Delta x_2}\sum_{u=1}^{p}\sum_{v=1}^{q}\rho_{u,v}e^{-i\mathbf{G^m\cdot x}}\Delta x_1\Delta x_2=\frac{1}{pq}\sum_{u=1}^{p}\sum_{v=1}^{q}\rho_{u,v}e^{-i2\pi\left(\frac{mu}{p}+\frac{nv}{q}\right)}\\
=\frac{1}{pq}\rho^{m,n}_{DFT},
\end{eqnarray}
where $\rho_{u,v}$ is the sampled density value, the components of $\mathbf{G^m}$ along the $\mathbf{e}_1$ and $\mathbf{e}_2$ directions are $(m\frac{2\pi}{p\Delta x_1}$, $n\frac{2\pi}{q\Delta x_2})$, and $\mathbf{x}=(u\Delta x_1,v\Delta x_2)$, $\rho^{m,n}_{DFT}$ is the Fourier coefficient calculated from either MATLAB's or NumPy's fft command. Similarly, the Fourier coefficients of compliance can be written as:
\begin{eqnarray}
\nonumber D^{m,n}=\frac{1}{p\Delta x_1 q\Delta x_2}\sum_{u=1}^{p}\sum_{v=1}^{q}D_{u,v}e^{-i\mathbf{G^m\cdot x}}\Delta x_1\Delta x_2=\frac{1}{pq}\sum_{u=1}^{p}\sum_{v=1}^{q}D_{u,v}e^{-i2\pi\left(\frac{mu}{p}+\frac{nv}{q}\right)}\\
=\frac{1}{pq}D^{m,n}_{DFT}.
\end{eqnarray}
The hexagonal unit cell made up of steel cylinders embedded in an epoxy matrix (Fig. \ref{fHex}a) is considered here. The diameter of the steel cylinders is 4mm and the lattice constant is 6.023mm. The material properties are taken from \citeN{vasseur2001experimental} and are reproduced here for reference

\begin{enumerate}
\item Steel: $C_{11}=264$ Gpa, $C_{44}=81$ Gpa, $\rho=7780$ kg/m$^3$
\item Epoxy: $C_{11}=7.54$ Gpa, $C_{44}=1.48$ Gpa, $\rho=1142$ kg/m$^3$
\end{enumerate}

\begin{figure}[htp]
\centering
\includegraphics[scale=.5]{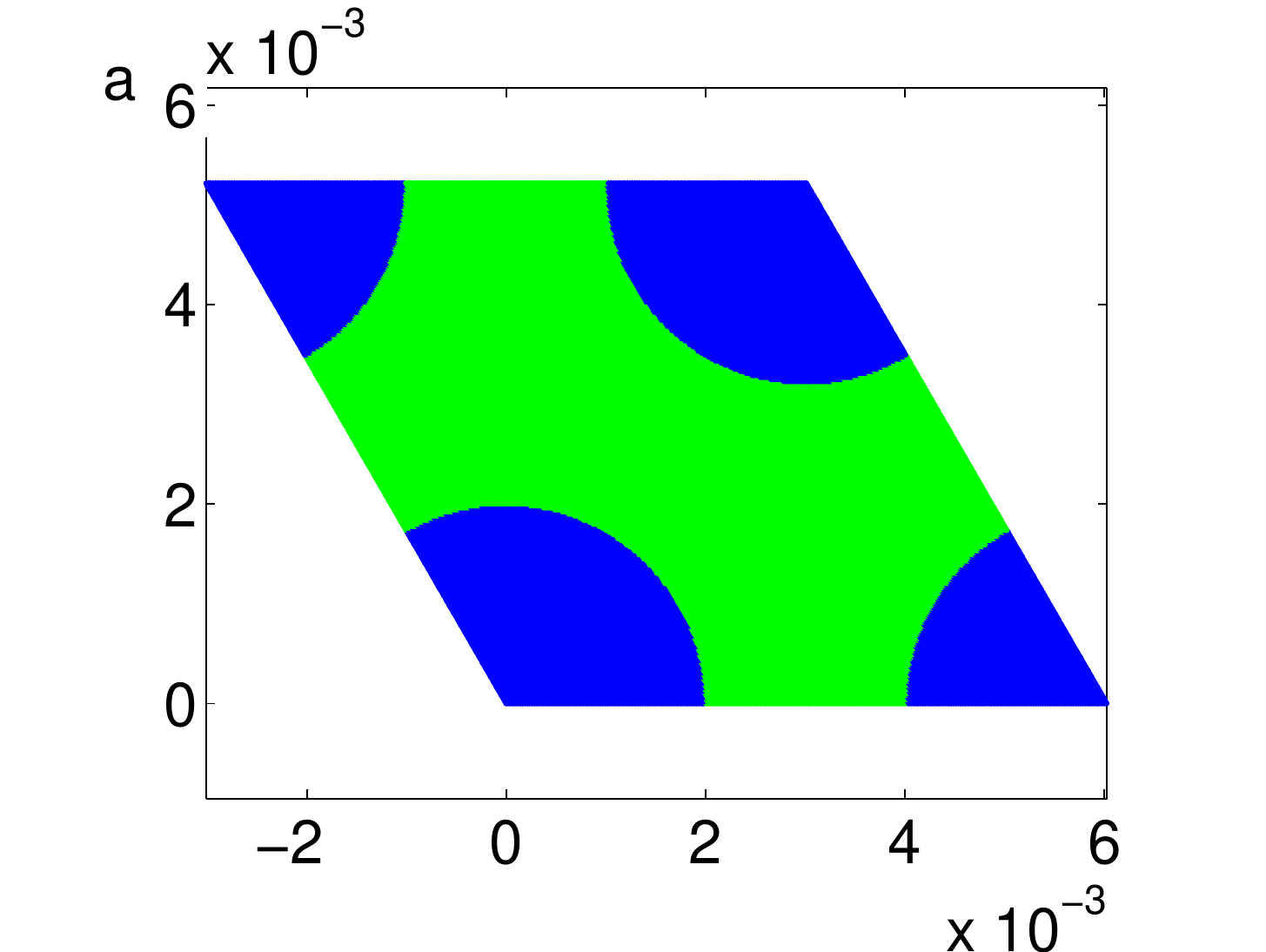}
\includegraphics[scale=.5]{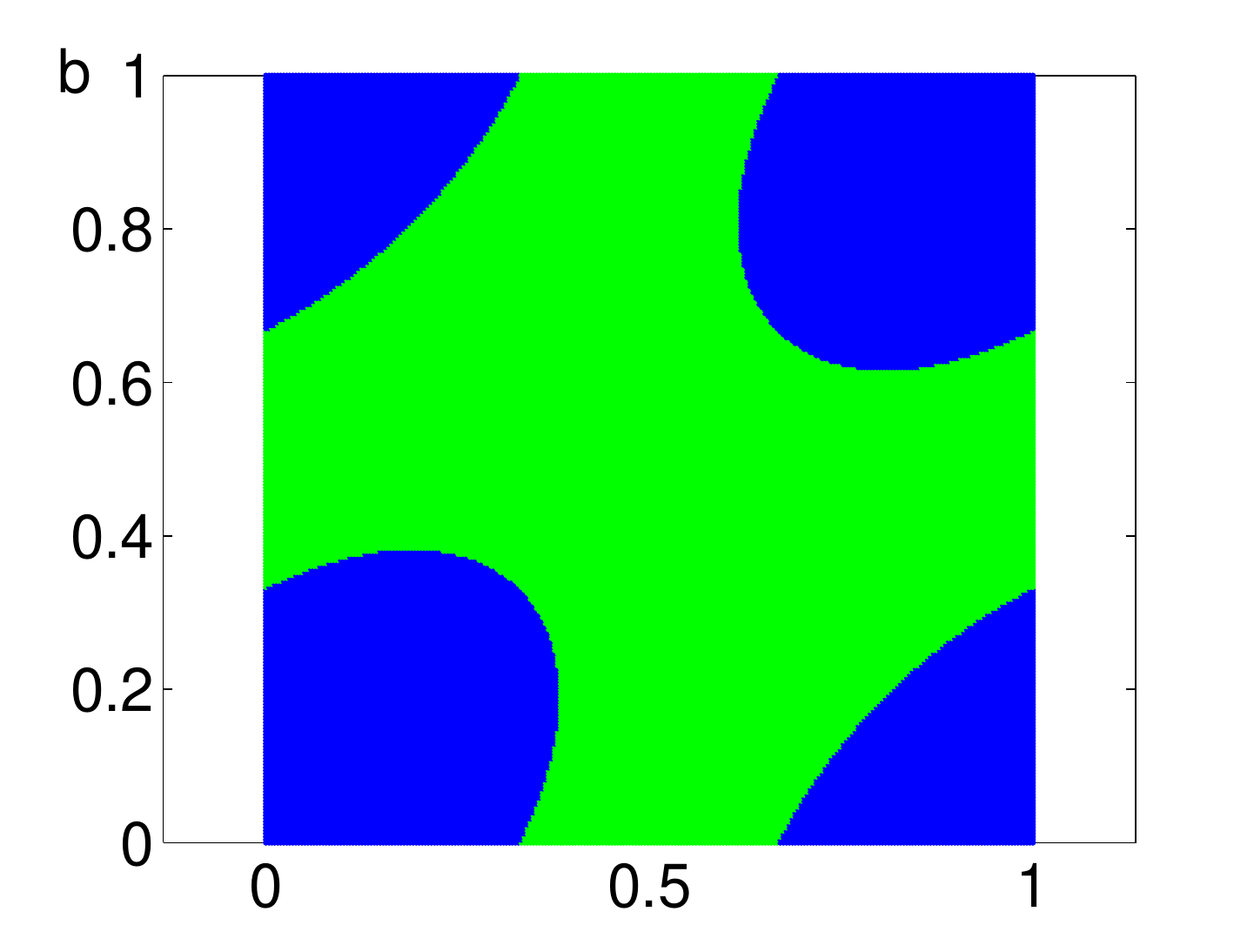}
\caption{a. Material properties are sampled over the unit cell, b. Material properties are projected onto the $(0,1)\times(0,1)$ domain.}\label{materialsampling2D}
\end{figure}
The material properties are sampled using $p=256$ and $q=256$ points along the $\mathbf{h}^1$ and $\mathbf{h}^2$ directions and then projected onto a $(0,1)\times(0,1)$ domain to calculate the Fourier coefficients. The material sampling is implemented using FEniCS \cite{logg2012automated} and the transformation is shown in Fig. \ref{materialsampling2D}.
 
\begin{figure}[htp]
\centering
\includegraphics[scale=.6]{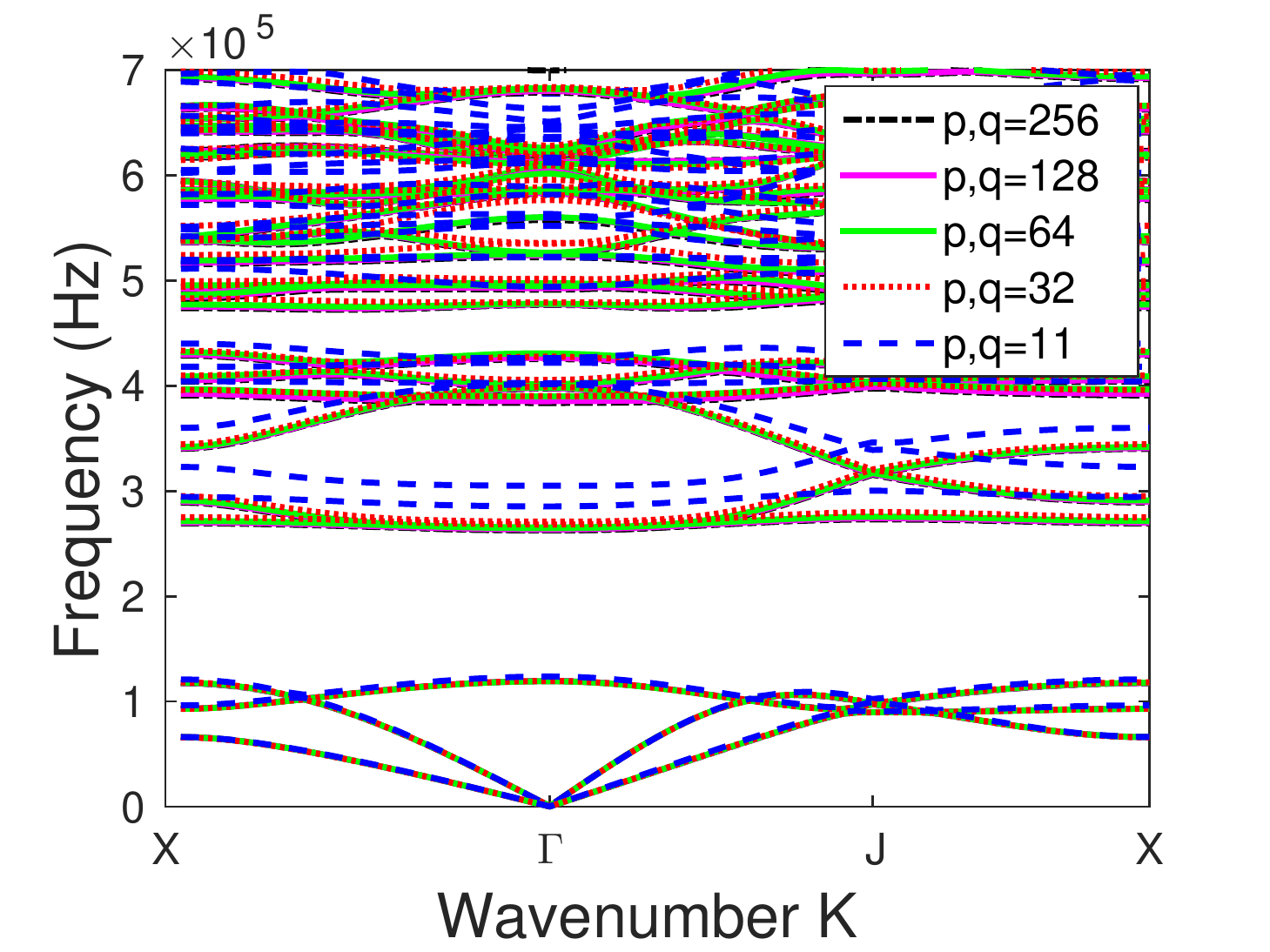}
\caption{Bandstructure calculation results using different number of material property sampling points.}\label{fCompare2D}
\end{figure}

The bandstructure is evaluated along the boundaries of the irreducible Brillouin zone. This boundary is denoted by the path $X-\Gamma-J-X$ and is shown in Fig. \ref{fHex}c in the reciprocal cell. We use a total of 242 terms in the expansion of field variables ($M=5$). This results in the simultaneous evaluation of the first 242 eigenvalues for each wavenumber point. The results in Fig. \ref{fCompare2D}, however, only show the lower frequency range. We note the existence of the all-angle stop-band for waves traveling in the plane of the unit cell in the frequency ranges of 120-262kHz and 427-473kHz. The locations of the stop-bands and the general shape of the pass-bands are shown to match very well with the results in \citeN{vasseur2001experimental} (Fig. 3 in that paper). It is similar to 1-D case, when the material sampling points along each direction are larger or equal to the number of coefficients ($M_p=11$ in this case) then the lower branches are always well estimated. For discretization higher than a certain value ($p,q\geq128$ in this case) the results for all the presently considered branches do not change appreciably (Fig. \ref{fCompare2D}).

\subsection{Example: face-centered cubic lattice}
\begin{figure}[htp]
\centering
\includegraphics[scale=.42]{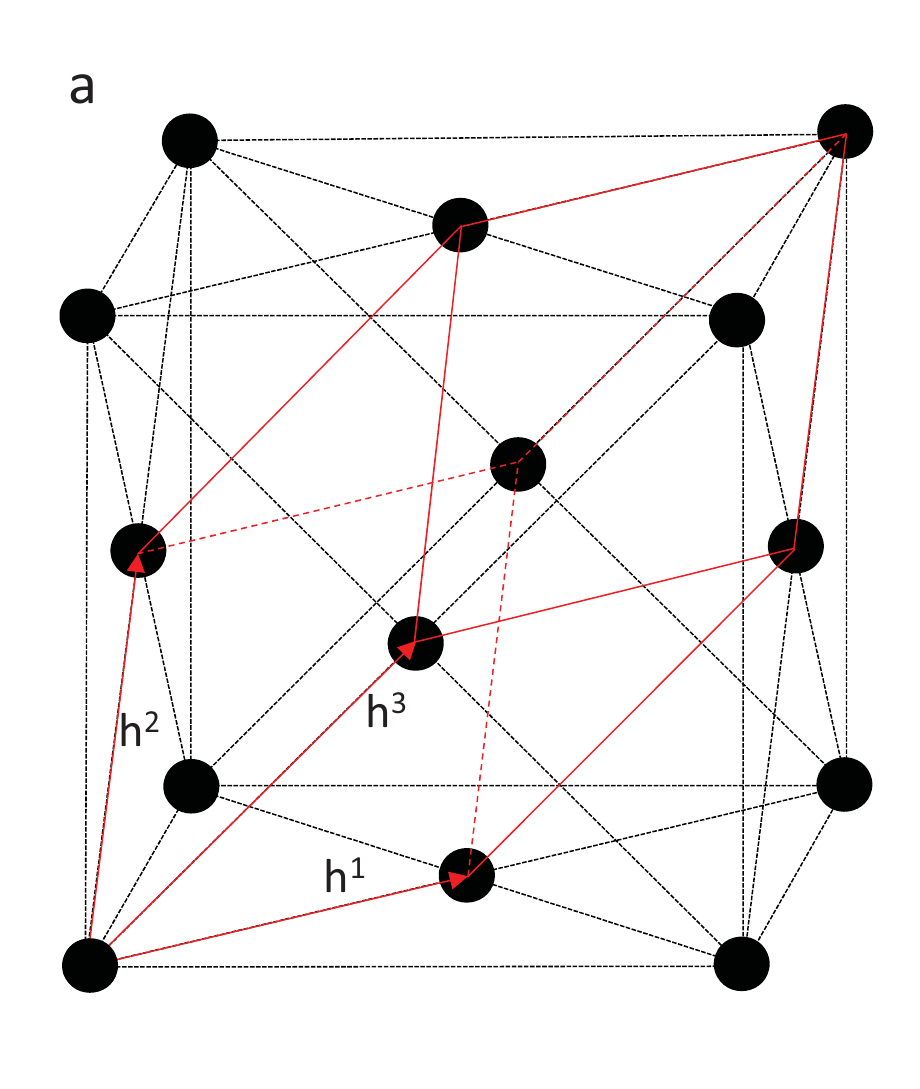}
\includegraphics[scale=.42]{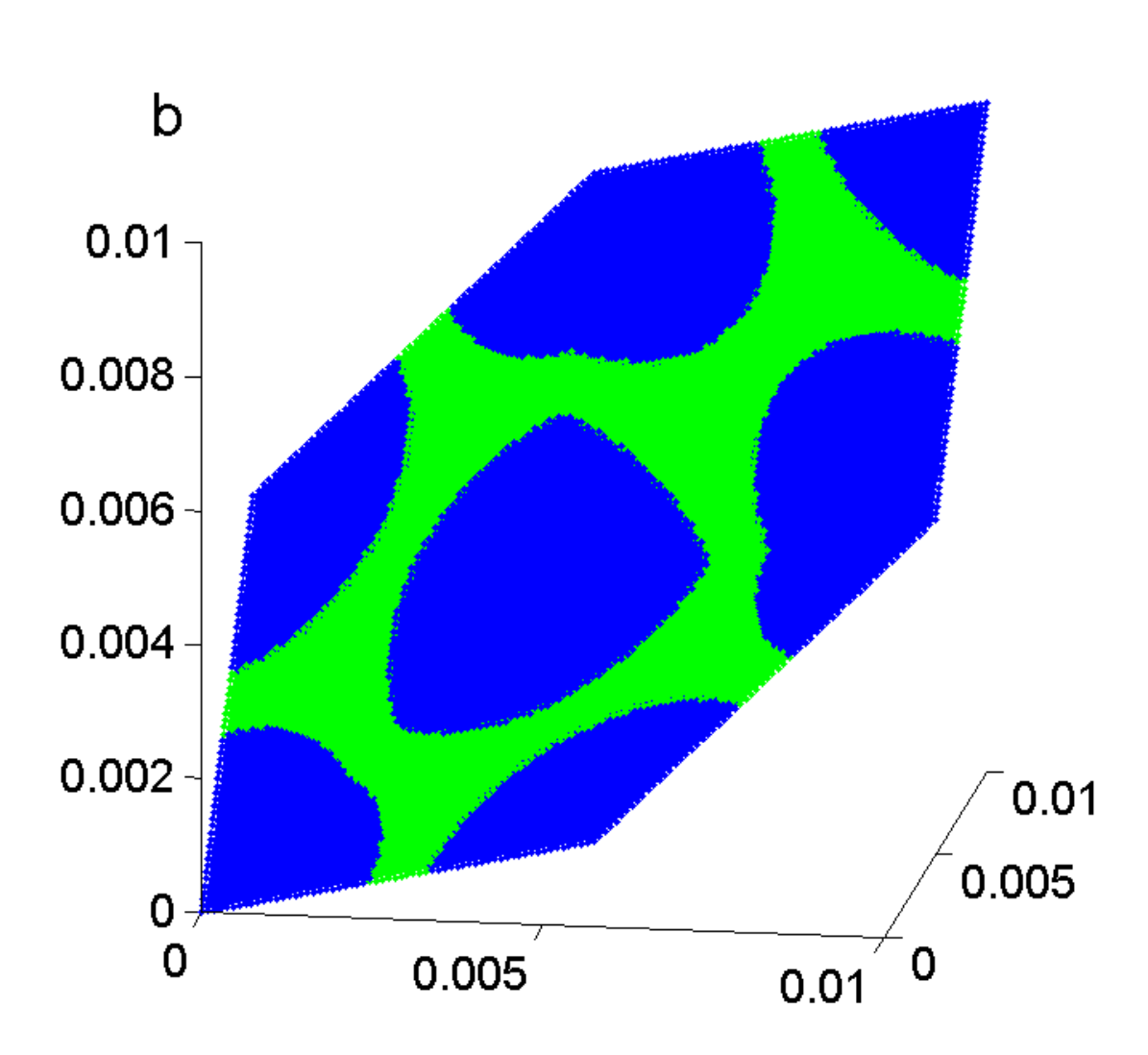}
\includegraphics[scale=.42]{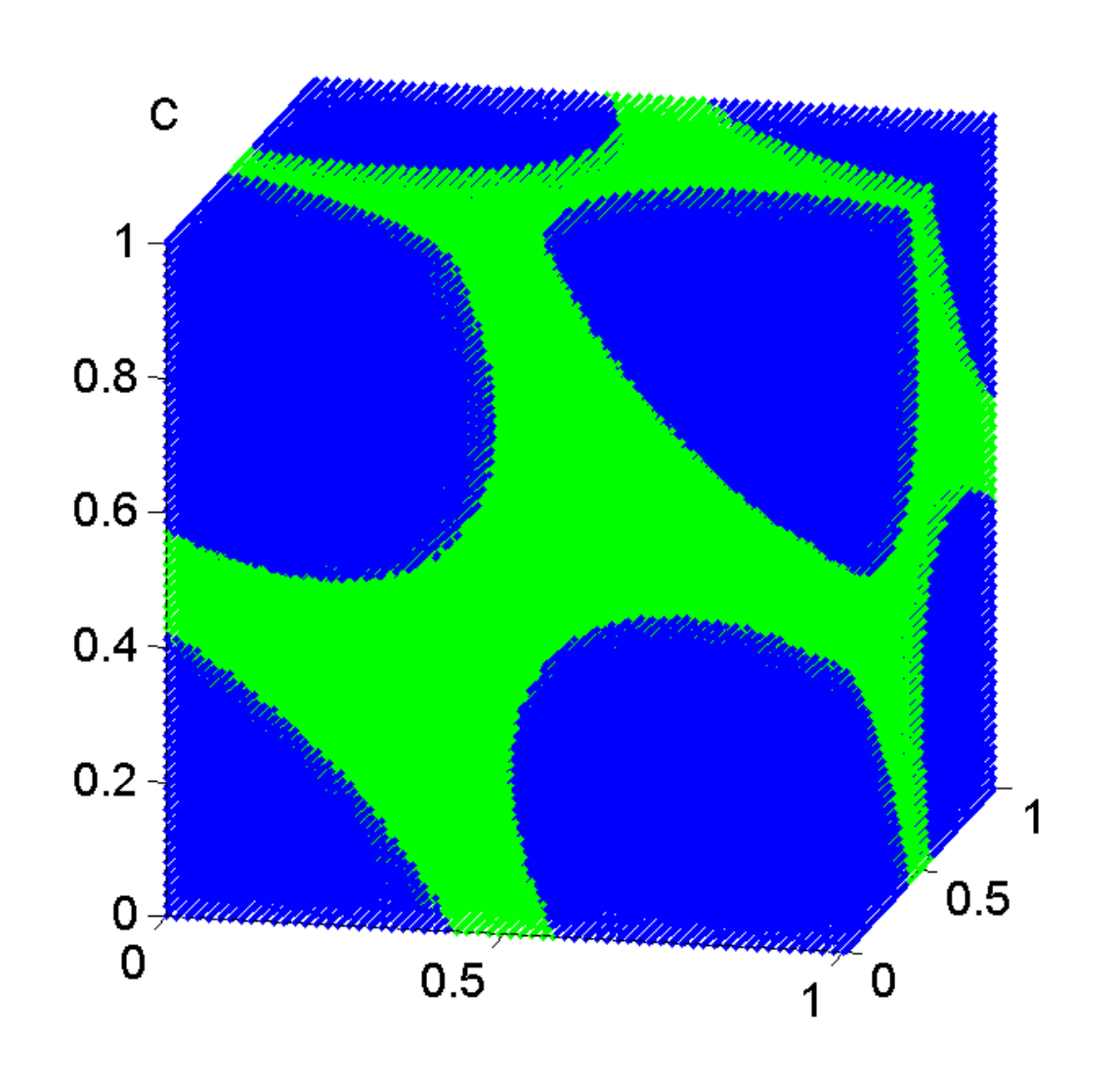}
\caption{a. Schematic of FCC Bravais lattice and its primitive cell, b. Material properties are sampled over the primitive cell, c. Material properties are projected onto the $(0,1)\times(0,1)\times(0,1)$ domain.}\label{materialsampling3D}
\end{figure}
In 3-D, the Fourier coefficients of density and compliance, given in Eq. (\ref{fouriercoef3D}), are in the form of three dimensional arrays. The material properties are sampled using $p$, $q$ and $r$ points along the $\mathbf{h}^1$, $\mathbf{h}^2$ and $\mathbf{h}^3$ directions, respectively. The $(l,m,n)$th Fourier coefficient of density and compliance can be obtained by using discrete Fourier transform as in the two dimensional case:
\begin{eqnarray}
\rho^{l,m,n}=\frac{1}{pqr}\sum_{u=1}^{p}\sum_{v=1}^{q}\sum_{w=1}^{r}\rho_{u,v,w}e^{-i2\pi\left(\frac{lu}{p}+\frac{mv}{q}+\frac{nw}{r}\right)}=\frac{1}{pqr}\rho^{l,m,n}_{DFT},\\
D^{l,m,n}=\frac{1}{pqr}\sum_{u=1}^{p}\sum_{v=1}^{q}\sum_{w=1}^{r}D_{u,v,w}e^{-i2\pi\left(\frac{lu}{p}+\frac{mv}{q}+\frac{nw}{r}\right)}=\frac{1}{pqr}D^{l,m,n}_{DFT}.
\end{eqnarray}
In the above equations, $\rho_{u,v,w}$ and $D_{u,v,w}$ are the sampled density and compliance values, and $\rho^{m,n}_{DFT}$ and $D^{l,m,n}_{DFT}$ are the Fourier coefficients calculated from either MATLAB's or NumPy's fft command.

\begin{figure}[htp]
\centering
\includegraphics[scale=.5]{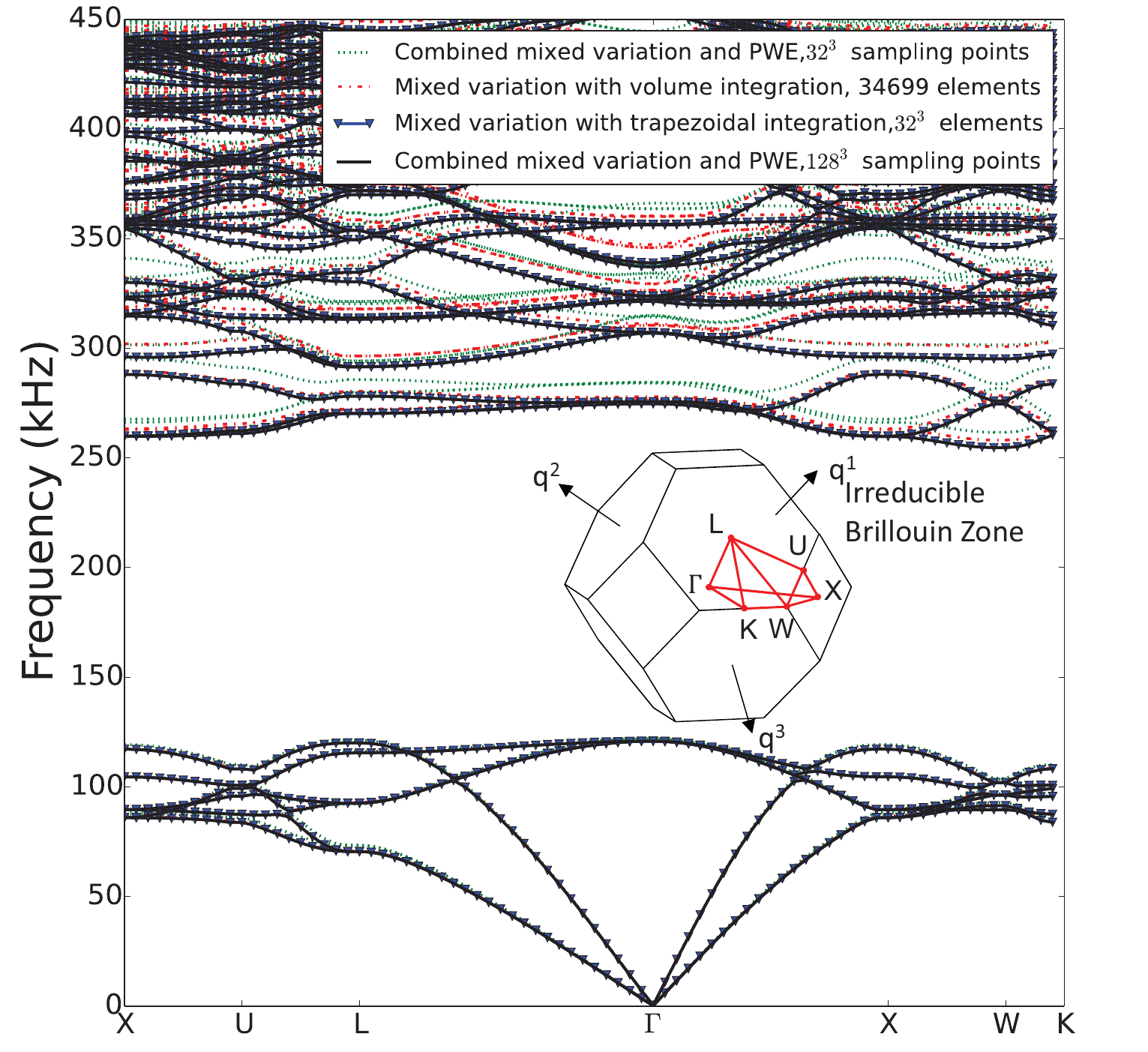}
\caption{Bandstructure calculated using $M=4$. For the current method we compare the results between $32^3$ and $128^3$ material sampling points, for mixed variation with volume integration 34699 unstructured tetrahedra elements are used and mixed variation with trapezoidal integrations $32^3$ structured cubic elements are used.}\label{fband3D}
\end{figure}

As an example, a face-centered cubic Bravais lattice (Fig. \ref{materialsampling3D}a) with a lattice constant of 1cm and a spherical inclusion of radius 3mm is considered. Material properties are the same as in the previous section. Properties are now sampled within the primitive cell using $p=q=r=128$ points along the $\mathbf{h}^1$, $\mathbf{h}^2$ and $\mathbf{h}^3$ directions and then projected onto a $(0,1)\times(0,1)\times(0,1)$ domain in order to calculate the Fourier coefficients. This transformation is shown in Figs. \ref{materialsampling3D}(b,c). The bandstructure is now evaluated using 2187 terms (M = 4) along the boundaries of irreducible Brillouin Zone (Fig. \ref{fband3D} inset). Fig. \ref{fband3D} (green dotted curve) shows the lower frequency range, where a complete bandgap starts from 121kHz to 256kHz. These results are in very good agreement with the FDTD calculations shown by \citeN{hsieh2006three}. Fig. \ref{fband3D} also shows the bandstructure results generated by using mixed variation with two different numerical integration methods, volume integration and trapezoidal integration. The dispersion relations agree very well on the first 6 bands, however, differences appear as frequency increases. The results generated using $128^3$ material sampling points agree very well with trapezoidal integration results (blue dashed). The material sampling time using FEnics for this 3-D case with $128^3$ material sampling points is only 158 seconds and it only needs to be carried out once. The matrix assembly time for $\mathbf{\Phi}$ and $\mathbf{\Omega}$, including FFT of the material properties is only 80 seconds for the $128^3$ points case comparing to about an hour when using trapezoidal integration over $32^3$ structure cubic elements. The convergence behavior of the three methods will be thoroughly discussed in the next section.

\section{Convergence and Stability}
It has been established through multiple lines of research that the mixed-variational method in general displays faster convergence than other comparable variational techniques and also the Plane Wave Expansion (PWE) method. For instance, theoretical arguments given by \citeN{babuska1978numerical} establish the superiority of the convergence characteristics of the mixed variational scheme over the Rayleigh quotient and the inverse Rayleigh quotient for eigenvalue calculations. Recently, we have also published a thorough numerical investigation of the convergence properties of the mixed variational method vis-a-vis the Rayleigh quotient, the displacement based Finite Element method, and the Plane Wave Expansion method \cite{lu2016variational}. In all the examples considered in the paper, the mixed variational method is seen to converge faster than the PWE method whose convergence properties are close to those of the Rayleigh quotient. It also exhibits faster convergence than the displacement based FE method. For details of the study we refer the reader to \citeN{lu2016variational}. Here a representative result is reproduced. 
\begin{figure}[htp]
\centering
\includegraphics[scale=.45]{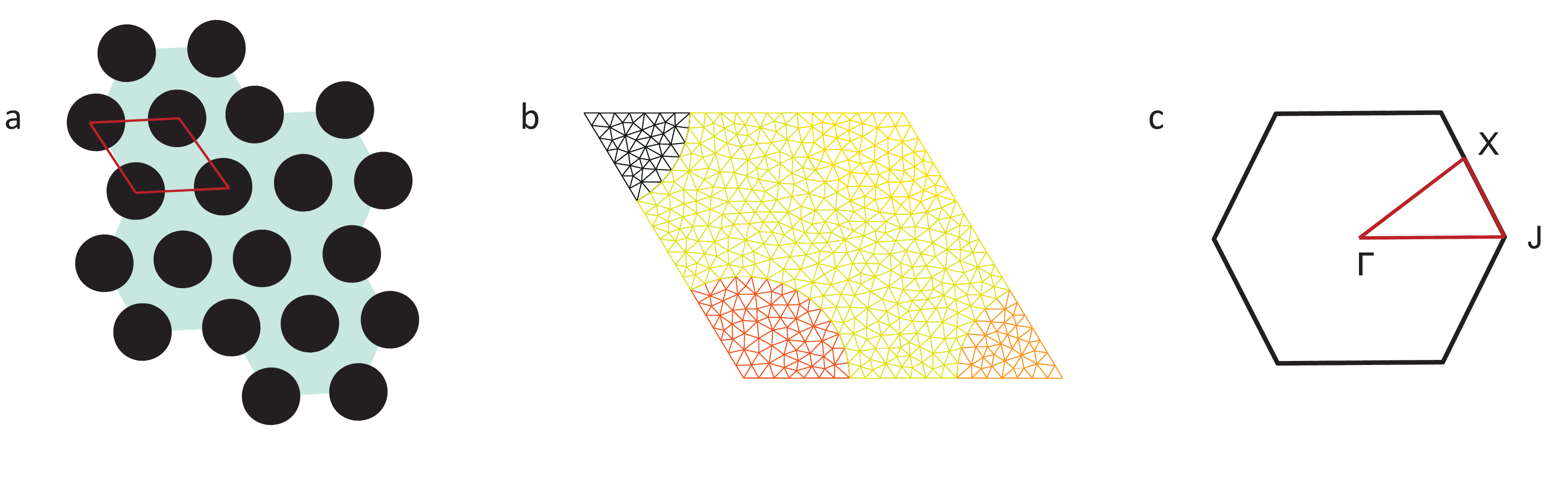}
\caption{2-D hexagonal phononic composite. a. Steel cylinders in Epoxy matrix, b. Finite element discretization of the unit cell, c. Irreducible Brillouin Zone.}\label{fHex}
\end{figure}
Fig. \ref{fHex} shows a 2-D hexagonal composite made of steel cylinders distributed in epoxy matrix. The error bound is given by the following inequality \cite{babuska1978numerical}:
\begin{equation}
\vert e \vert=\vert \lambda_0 - \lambda_0^M \vert \leq CM^{-\xi},
\end{equation}
where $\vert e \vert$ is the absolute error, $\lambda_0$ is the exact eigenvalue, $\lambda_0^M$ is the eigenvalue approximated by $M$ Fourier terms, $C$ is a constant and $\xi$ is the convergence rate. Convergence rate can, therefore, be approximated by the slope of the following:
\begin{equation}
\log \vert e \vert \leq -\xi \log M +\log C,
\end{equation}
where $C$ is the relative error when $M=1$ and $\xi$ is the convergence rate. Table \ref{averageconvr2D} provides comparisons between the average convergence rates of mixed-variational method and the PWE method. The convergence rates are shown for the first four phononic curves and at 4 different wave-vector points. At all points of computations the mixed method converges significantly faster than the plane wave method. Since the accuracy of the solution depends upon the convergence rate \cite{babuska1978numerical,lu2016variational} it is clear that the mixed method can be used to solve for larger number of phononic bands with more accuracy than the Plane Wave method for similar sizes of the eigenvalue matrix. 
\begin{table}[htp]
\caption{Average convergence rates for the computation of eigenvalues at $X$ and the midpoints of $X-\Gamma$, $\Gamma-J$, $J-X$. $\xi_m,\xi_p$ refer to the convergence rates of the mixed method and PWE, respectively. They are the slopes of the linear relationship between the natural $\log$ of relative errors of frequencies and $\log M$. }\label{averageconvr2D}
\centering
\begin{tabular}{p{1.2cm}p{1.2cm}p{1.2cm}p{1.2cm}p{1.2cm}p{1.2cm}p{1.2cm}p{1.2cm}l}
\hline & \multicolumn{2}{c}{$Curve_1$}  & \multicolumn{2}{c}{$Curve_2$} & \multicolumn{2}{c}{$Curve_3$} & \multicolumn{2}{c}{$Curve_4$}\\
\cline{2-3}  \cline{4-5}  \cline{6-7}  \cline{8-9}
& $\xi_{m}$ & $\xi_{p}$  & $\xi_{m}$ & $\xi_{p}$  & $\xi_{m}$ & $\xi_{p}$ &  $\xi_{m}$ & $\xi_{p}$\\
\hline X  & 2.0434 & 1.2948 & 2.1401 & 1.2884 & 2.2317 & 1.2074 & 2.6812 & 1.1971\\
\hline X-$\Gamma$ & 1.9384 & 1.2539  & 1.8682 & 1.2209  & 2.3124 & 1.2968 & 2.4236 & 1.2325\\
\hline $\Gamma$-J & 1.9363 & 1.2651 & 1.7762 & 1.1713 & 2.2346 & 1.3046 & 2.4417 & 1.2349\\
\hline J-X & 2.0715 & 1.2647 & 2.1718 & 1.2690 & 2.1950 & 1.2336 & 2.6764 & 1.1817\\
\hline
\end{tabular}
\end{table}

In the above calculations for the mixed method, the integrals occurring in Eq. (\ref{equationshomogeneousMatrix}) are numerically calculated over $\Omega$. Numerical integration is achieved by first dividing the domain $\Omega$ into $P$ subdomains $\Omega_i,i=1,2...P$ through a freely available open source Finite Element software \cite{geuzaine2009gmsh}. The volume integral of any function $F(\mathbf{x})$ is then approximated as:
\begin{equation}\label{eConvInt}
\int_\Omega F(\mathbf{x})d\Omega=\sum_i^PF_iV_i,
\end{equation}
where $F_i$ is the value of the function $F(\mathbf{x})$ evaluated at the centroid of $\Omega_i$ and $V_i$ is the volume of $\Omega_i$. MATLAB (or Python) routines are then used to calculate the volumes and centroids of these subdomains and the required integrals over $\Omega$. In our case, the integrands are of the form $F(\mathbf{x})=\mu(\mathbf{x})\exp(i\mathbf{G^n}\cdot\mathbf{x})$ where $\mu(\mathbf{x})$ is some spatially dependent material property (density or elasticitiy tensor component) and $\mathbf{G^n}=n_i\mathbf{q}_i$. The largest and smallest (largest negative) values that $n_i$ can take depend upon how many trigonometric terms are used in the expansion of the displacement and stress fields and they correspond to the smallest wavelengths used in the expansion. The number of terms also directly determines the size of the eigenvalue problem and the number of eigenvalues that can be determined from its solution. In phononic applications there is an incentive to increase the number of terms in the trigonometric expansions not only in order to calculate more eigenvalues but also to, hopefully, calculate the smaller eigenvalues with higher accuracies. However, if the integrals are calculated as in Eq. (\ref{eConvInt}) then the level of discretization (size of the smallest element) of the domain can be chosen independently of the smallest wavelength which exists in the expansion. This leads to computational stability issues which are neatly avoided if the material properties are themselves expanded in Fourier series as shown in this paper. This is due to the fact that the DFT coefficients, which our current approach is based upon, automatically connects the space discretization with the smallest wavelength existing in the expansion through the Nyquist relationship \cite{nyquist1928certain}. FFT algorithms ensure that the DFT coefficients corresponding to this smallest wavelength can be calculated with high accuracy and without any aliasing errors \cite{cooley1965algorithm}. However, this naturally raises another question: whether the stability issue will be addressed if a better numerical integration scheme is applied. 

\begin{figure}[htp]
\centering
\includegraphics[scale=.5]{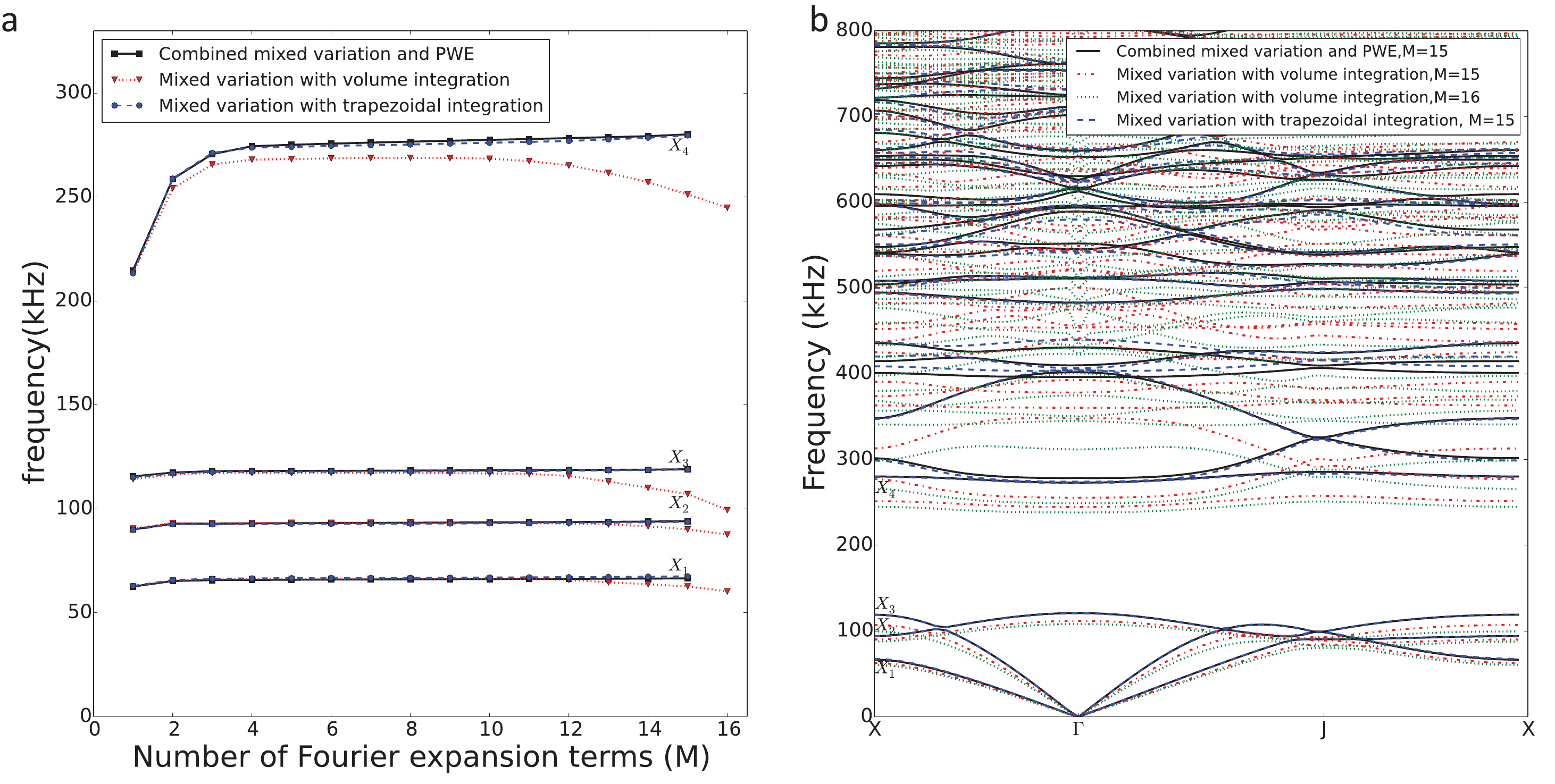}
\caption{a. Bandstructure calculated by two methods using $M=15$. b. Frequency values on the first four curves at $X$ for different $M$. $32^2$ material sampling points are used in combined mixed variation and PWE, $1112$ unstructured triangular mesh elements are used in mixed variation with volume integrations and $32^2$ structured quad elements are used in mixed variation with trapezoidal integrations.}\label{convergence}
\end{figure}

First, to elaborate upon the stability issue consider again the solution to the phononic problem shown in Fig. \ref{fHex}. The mixed-variational method using volume integration Eq. (\ref{eConvInt}) is used to calculate the corresponding eigenvalue problem. This is done by discretizing its unit cell into 1112 triangular mesh elements. The red curve in Fig. (\ref{convergence}a) shows the first four frequencies as a function of the number of Fourier expansion terms used in each direction $M$. In the next step we calulate the bandstructure of the same phononic crystal using mixed variation with a higher order numerical integration scheme (trapezoidal integration). In 1-D it is given by: 
\begin{equation}\label{eConvInttrapz}
\int_a^b f(x)dx=\dfrac{b-a}{2N}\sum_{n=1}^N(f(x_n)+f(x_{n+1})),
\end{equation}
where $N$ is the number of elements. Results are shown as blue curves in Fig. (\ref{convergence}a). These are compared with the mixed-variation calculations which employ the Fourier coefficients of the material properties and avoid direct numerical integration over a Finite Element mesh (black curve in Fig. \ref{convergence}a). For smaller values of $M$, the three methods give essentially the same results, especially for the first three branches ($X_1$ to $X_3$.) However, as higher values of $M$ are considered, the results of the calculations based upon direct volume integration worsen but the higher order integration method and the current method give similar and stable results. Furthermore for a given space discretization, the current method automatically places an upper limit of the highest $M$ value which may used accurately. However, numerical integration method (centroid based or higher order) places no such limit. In the present example, for instance, $M=16$ could be used with the formulation based upon numerical integration and it results in significantly worse accuracy (Fig. \ref{convergence}a). This stability effect is further elaborated upon in the full bandstructure calculation ( Fig. \ref{convergence}b). In these calculations we have used $\approx 1100$ elements in spatial discretization for the thee methods. An $M=15$ calculation through the current scheme (black curve) results in a bandstructure which shows high fidelity when compared with Fig. \ref{fCompare2D}. It should be noted that these calculations have been carried out on a relatively coarse grid ($\approx 32\times 32$). At the same discretization level and the same number of Fourier expansion terms ($M=15$), the volume integration method based upon centroidal integration (red dashed) shows significant deviation from the black curve even at the lower branches. The stability issues of this method really become clear when ($M=16$) is considered for volume integration calculations (green dotted). This results in anomalous sharp dips in the bandstructure at the $\Gamma$ point beginning from the $12^\mathrm{th}$ branch ($\approx 400 \text{kHz}$). For the same $M$ value ($M=15$), higher order numerical integration (blue dashed) performs better and shows close correspondence with the Fourier results for the lower branches. We note that this improvement in performance is achieved, however, at a higher computational cost. Furthermore, the higher order integration results in Fig. (\ref{convergence}b) start to deviate from the results of the current method at as low as the $5^{th}$ branch. 

Similar performance is seen in the 3-D case. For the same $M$ value ($M=4$), higher order numerical integration (blue solid marked by triangles) performs better and shows close corresondence with the Fourier results over finer mesh. This improvements is also achieved at a higher computational cost. Lower order integration start to deviate from the results of the current method at as low as the $10^{th}$ branch. The Fourier results generated over coarse mesh deviates over the $7^{th}$ bands. Finer mesh should be employed for 3D case if using the current method but it involves significantly less computation effort comparing to higher order numerical integration. 

In summary, the above calculations show that the mixed-variation method, based upon the Fourier decomposition of material properties, shows convergence parity with formulations based upon numerical integration (centroid based or higher order) when $M$ values are low (for a given spatial disretization level). However the Fourier decomposition technique renders the method stable over the entire range of $M$ values as allowed by the discretization level. Within this range it outperforms centroid based numerical integration which shows signs of instability at higher $M$ values. Higher order numerical integration is stable and its results are close to those obtained by Fourier decomposition. However, this comes at a higher computational cost compared with centroid based numerical integration which is already significantly slower than the Fourier decomposition method (Fig. \ref{fCompTime}).

\section{Conclusions}

In this paper we have proposed a combination of the salient features of the PWE method (Fourier expansion of material properties) and the mixed variational method. The former is easy to implement while the latter has been shown to have superior convergence property with respect to PWE and displacement based variational methods. The resulting scheme obviates the need for numerical integration which originally existed in the variational method. Furthermore, it allows us to express all the relevant matrix elements through closed form expressions depending directly upon the discrete Fourier Transform (DFT) components. The minimum number of Fourier coefficients necessary for the matrix assembly is determined by the number of Fourier expansion terms for field variables. Although the accuracy of Fourier coefficients depends on the sampling resolution, increasing the resolution has minimum effect on improving the accuracy of solution beyond certain limit. The DFT computation time is negligible compared to numerical integration over the same number of elements, thus resulting in further accelerated matrix assembly. This results in a fast and efficient phononic eigenvalue algorithm which also possesses easy implementation similar to PWE. Comparative examples in this paper show good agreement with reference results. The convergence and stability study shows that the present method is always stable over the entire range of the expansion terms as allowed by the spatial discretization when compared with the centroid based zero order numerical integration scheme. The higher order numerical integration comes to the accuracy of the present method but with significantly more computational expense.

\section{Acknowledgment}

A.S. acknowledges support from the NSF CAREER grant $\#$ 1554033 to the Illinois Institute of Technology.


\begin{thebibliography}{}

\bibitem[\protect\citeauthoryear{}{Amirkulova and
  Norris}{2015}]{amirkulova2015acoustic}
Amirkulova, F.~A. and Norris, A.~N. (2015).
\newblock ``Acoustic multiple scattering using recursive algorithms.''\ {\em
  Journal of Computational Physics}, 299, 787--803.

\bibitem[\protect\citeauthoryear{}{Babu{\v{s}}ka and
  Osborn}{1978}]{babuska1978numerical}
Babu{\v{s}}ka, I. and Osborn, J.~E. (1978).
\newblock ``Numerical treatment of eigenvalue problems for differential
  equations with discontinuous coefficients.''\ {\em Mathematics of
  Computation}, 32(144), 991--1023.

\bibitem[\protect\citeauthoryear{}{Bao et~al.\@}{2011}]{bao2011dynamic}
Bao, J., Shi, Z., and Xiang, H. (2011).
\newblock ``Dynamic responses of a structure with periodic foundations.''\ {\em
  Journal of Engineering Mechanics}, 138(7), 761--769.

\bibitem[\protect\citeauthoryear{}{Bilal and
  Hussein}{2011}]{bilal2011ultrawide}
Bilal, O.~R. and Hussein, M.~I. (2011).
\newblock ``Ultrawide phononic band gap for combined in-plane and out-of-plane
  waves.''\ {\em Physical Review E}, 84(6), 065701.

\bibitem[\protect\citeauthoryear{}{Cao et~al.\@}{2004}]{cao2004finite}
Cao, Y., Hou, Z., and Liu, Y. (2004).
\newblock ``Finite difference time domain method for band-structure
  calculations of two-dimensional phononic crystals.''\ {\em Solid state
  communications}, 132(8), 539--543.

\bibitem[\protect\citeauthoryear{}{Cervera
  et~al.\@}{2001}]{cervera2001refractive}
Cervera, F., Sanchis, L., S{\'a}nchez-P{\'e}rez, J.~V., Martinez-Sala, R.,
  Rubio, C., Meseguer, F., L{\'o}pez, C., Caballero, D., and
  S{\'a}nchez-Dehesa, J. (2001).
\newblock ``Refractive acoustic devices for airborne sound.''\ {\em Physical
  review letters}, 88(2), 023902.

\bibitem[\protect\citeauthoryear{}{Chaunsali
  et~al.\@}{2016}]{chaunsali2016stress}
Chaunsali, R., Li, F., and Yang, J. (2016).
\newblock ``Stress wave isolation by purely mechanical topological phononic
  crystals.''\ {\em Scientific Reports}, 6.

\bibitem[\protect\citeauthoryear{}{Cleland et~al.\@}{2001}]{cleland2001thermal}
Cleland, A.~N., Schmidt, D.~R., and Yung, C.~S. (2001).
\newblock ``Thermal conductance of nanostructured phononic crystals.''\ {\em
  Physical Review B}, 64(17), 172301.

\bibitem[\protect\citeauthoryear{}{Cooley and
  Tukey}{1965}]{cooley1965algorithm}
Cooley, J.~W. and Tukey, J.~W. (1965).
\newblock ``An algorithm for the machine calculation of complex fourier
  series.''\ {\em Mathematics of computation}, 19(90), 297--301.

\bibitem[\protect\citeauthoryear{}{Dertimanis
  et~al.\@}{2016}]{dertimanis2016feasibility}
Dertimanis, V.~K., Antoniadis, I.~A., and Chatzi, E.~N. (2016).
\newblock ``Feasibility analysis on the attenuation of strong ground motions
  using finite periodic lattices of mass-in-mass barriers.''\ {\em Journal of
  Engineering Mechanics},  04016060.

\bibitem[\protect\citeauthoryear{}{Geuzaine and
  Remacle}{2009}]{geuzaine2009gmsh}
Geuzaine, C. and Remacle, J.-F. (2009).
\newblock ``Gmsh: A 3-d finite element mesh generator with built-in pre-and
  post-processing facilities.''\ {\em International Journal for Numerical
  Methods in Engineering}, 79(11), 1309--1331.

\bibitem[\protect\citeauthoryear{}{Haque and Shim}{2016}]{haque2016spatial}
Haque, A.~T. and Shim, J. (2016).
\newblock ``On spatial aliasing in the phononic band-structure of layered
  composites.''\ {\em International Journal of Solids and Structures}.

\bibitem[\protect\citeauthoryear{}{Hladky-Hennion and
  Decarpigny}{1991}]{hladky1991analysis}
Hladky-Hennion, A.-C. and Decarpigny, J.-N. (1991).
\newblock ``Analysis of the scattering of a plane acoustic wave by a doubly
  periodic structure using the finite element method: Application to alberich
  anechoic coatings.''\ {\em The Journal of the Acoustical Society of America},
  90(6), 3356--3367.

\bibitem[\protect\citeauthoryear{}{Hsieh et~al.\@}{2006}]{hsieh2006three}
Hsieh, P.-F., Wu, T.-T., and Sun, J.-H. (2006).
\newblock ``Three-dimensional phononic band gap calculations using the fdtd
  method and a pc cluster system.''\ {\em Ultrasonics, Ferroelectrics, and
  Frequency Control, IEEE Transactions on}, 53(1), 148--158.

\bibitem[\protect\citeauthoryear{}{Hussein et~al.\@}{2015}]{hussein2015flow}
Hussein, M., Biringen, S., Bilal, O., and Kucala, A. (2015).
\newblock ``Flow stabilization by subsurface phonons.''\ {\em Proc. R. Soc. A},
  Vol. 471, The Royal Society,  20140928.

\bibitem[\protect\citeauthoryear{}{Hussein
  et~al.\@}{2014}]{hussein2014dynamics}
Hussein, M.~I., Leamy, M.~J., and Ruzzene, M. (2014).
\newblock ``Dynamics of phononic materials and structures: Historical origins,
  recent progress, and future outlook.''\ {\em Applied Mechanics Reviews},
  66(4), 040802.

\bibitem[\protect\citeauthoryear{}{Kafesaki and
  Economou}{1999}]{kafesaki1999multiple}
Kafesaki, M. and Economou, E. (1999).
\newblock ``Multiple-scattering theory for three-dimensional periodic acoustic
  composites.''\ {\em Physical Review B}, 60(17), 11993--12001.

\bibitem[\protect\citeauthoryear{}{Khelif et~al.\@}{2003}]{khelif2003trapping}
Khelif, A., Choujaa, A., Djafari-Rouhani, B., Wilm, M., Ballandras, S., and
  Laude, V. (2003).
\newblock ``Trapping and guiding of acoustic waves by defect modes in a
  full-band-gap ultrasonic crystal.''\ {\em physical Review B}, 68(21), 214301.

\bibitem[\protect\citeauthoryear{}{Kushwaha
  et~al.\@}{1993}]{kushwaha1993acoustic}
Kushwaha, M., Halevi, P., Dobrzynski, L., and Djafari-Rouhani, B. (1993).
\newblock ``Acoustic band structure of periodic elastic composites.''\ {\em
  Physical Review Letters}, 71(13), 2022--2025.

\bibitem[\protect\citeauthoryear{}{Kushwaha
  et~al.\@}{1994}]{kushwaha1994theory}
Kushwaha, M., Halevi, P., Martinez, G., Dobrzynski, L., and Djafari-Rouhani, B.
  (1994).
\newblock ``Theory of acoustic band structure of periodic elastic
  composites.''\ {\em Physical Review B}, 49(4), 2313--2322.

\bibitem[\protect\citeauthoryear{}{Landry et~al.\@}{2008}]{landry2008complex}
Landry, E.~S., Hussein, M.~I., and McGaughey, A. J.~H. (2008).
\newblock ``Complex superlattice unit cell designs for reduced thermal
  conductivity.''\ {\em Physical Review B}, 77(18), 184302.

\bibitem[\protect\citeauthoryear{}{Li et~al.\@}{2011}]{li2011tunable}
Li, X.-F., Ni, X., Feng, L., Lu, M.-H., He, C., and Chen, Y.-F. (2011).
\newblock ``Tunable unidirectional sound propagation through a
  sonic-crystal-based acoustic diode.''\ {\em Physical review letters}, 106(8),
  084301.

\bibitem[\protect\citeauthoryear{}{Logg et~al.\@}{2012}]{logg2012automated}
Logg, A., Mardal, K.-A., and Wells, G. (2012).
\newblock {\em Automated solution of differential equations by the finite
  element method: The FEniCS book}, Vol.~84.
\newblock Springer Science \& Business Media.

\bibitem[\protect\citeauthoryear{}{Lu and Srivastava}{2016}]{lu2016variational}
Lu, Y. and Srivastava, A. (2016).
\newblock ``Variational methods for phononic calculations.''\ {\em Wave
  Motion}, 60, 46--61.

\bibitem[\protect\citeauthoryear{}{Lu et~al.\@}{2017}]{lu20173}
Lu, Y., Yang, Y., Guest, J.~K., and Srivastava, A. (2017).
\newblock ``3-d phononic crystals with ultra-wide band gaps.''\ {\em Scientific
  Reports}, 7(43407).

\bibitem[\protect\citeauthoryear{}{Minagawa and
  Nemat-Nasser}{1976}]{minagawa1976harmonic}
Minagawa, S. and Nemat-Nasser, S. (1976).
\newblock ``Harmonic waves in three-dimensional elastic composites.''\ {\em
  International Journal of Solids and Structures}, 12(11), 769--777.

\bibitem[\protect\citeauthoryear{}{Mitchell
  et~al.\@}{2016}]{mitchell2016effect}
Mitchell, S.~J., Pandolfi, A., and Ortiz, M. (2016).
\newblock ``Effect of brittle fracture in a metaconcrete slab under shock
  loading.''\ {\em Journal of Engineering Mechanics}, 142(4), 04016010.

\bibitem[\protect\citeauthoryear{}{Nemat-Nasser and
  Srivastava}{2011}]{nemat2011overall}
Nemat-Nasser, S. and Srivastava, A. (2011).
\newblock ``Overall dynamic constitutive relations of layered elastic
  composites.''\ {\em Journal of the Mechanics and Physics of Solids}, 59(10),
  1953--1965.

\bibitem[\protect\citeauthoryear{}{Nemat-Nasser
  et~al.\@}{2011}]{nemat2011homogenization}
Nemat-Nasser, S., Willis, J.~R., Srivastava, A., and Amirkhizi, A.~V. (2011).
\newblock ``Homogenization of periodic elastic composites and locally resonant
  sonic materials.''\ {\em Physical Review B}, 83(10), 104103.

\bibitem[\protect\citeauthoryear{}{Nyquist}{1928}]{nyquist1928certain}
Nyquist, H. (1928).
\newblock ``Certain topics in telegraph transmission theory.''\ {\em
  Transactions of the American Institute of Electrical Engineers}, 47(2),
  617--644.

\bibitem[\protect\citeauthoryear{}{Sigalas and
  Economou}{1993}]{sigalas1993band}
Sigalas, M. and Economou, E. (1993).
\newblock ``Band structure of elastic waves in two dimensional systems.''\ {\em
  Solid State Communications}, 86(3), 141--143.

\bibitem[\protect\citeauthoryear{}{Sigalas
  et~al.\@}{2000}]{sigalas2000theoretical}
Sigalas, M. et~al.\@ (2000).
\newblock ``Theoretical study of three dimensional elastic band gaps with the
  finite-difference time-domain method.''\ {\em Journal of Applied Physics},
  87(6), 3122--3125.

\bibitem[\protect\citeauthoryear{}{Srivastava}{2015}]{srivastava2015elastic}
Srivastava, A. (2015).
\newblock ``Elastic metamaterials and dynamic homogenization: a review.''\ {\em
  International Journal of Smart and Nano Materials}, 6(1), 41--60.

\bibitem[\protect\citeauthoryear{}{Srivastava}{2016}]{srivastava2016metamaterial}
Srivastava, A. (2016).
\newblock ``Metamaterial properties of periodic laminates.''\ {\em Journal of
  the Mechanics and Physics of Solids}, 96, 252--263.

\bibitem[\protect\citeauthoryear{}{Srivastava and
  Nemat-Nasser}{2012}]{srivastava2012overall}
Srivastava, A. and Nemat-Nasser, S. (2012).
\newblock ``Overall dynamic properties of three-dimensional periodic elastic
  composites.''\ {\em Proc. R. Soc. A}, Vol. 468, The Royal Society,  269--287.

\bibitem[\protect\citeauthoryear{}{Srivastava and
  Nemat-Nasser}{2014}]{srivastava2014mixed}
Srivastava, A. and Nemat-Nasser, S. (2014).
\newblock ``Mixed-variational formulation for phononic band-structure
  calculation of arbitrary unit cells.''\ {\em Mechanics of Materials}, 74,
  67--75.

\bibitem[\protect\citeauthoryear{}{Sukhovich
  et~al.\@}{2008}]{sukhovich2008negative}
Sukhovich, A., Jing, L., and Page, J.~H. (2008).
\newblock ``Negative refraction and focusing of ultrasound in two-dimensional
  phononic crystals.''\ {\em Physical Review B}, 77(1), 014301.

\bibitem[\protect\citeauthoryear{}{Vasseur
  et~al.\@}{1994}]{vasseur1994complete}
Vasseur, J., Djafari-Rouhani, B., Dobrzynski, L., Kushwaha, M., and Halevi, P.
  (1994).
\newblock ``Complete acoustic band gaps in periodic fibre reinforced composite
  materials: the carbon/epoxy composite and some metallic systems.''\ {\em
  Journal of Physics: Condensed Matter}, 6(42), 8759.

\bibitem[\protect\citeauthoryear{}{Vasseur
  et~al.\@}{2001}]{vasseur2001experimental}
Vasseur, J.~O., Deymier, P.~A., Chenni, B., Djafari-Rouhani, B., Dobrzynski,
  L., and Prevost, D. (2001).
\newblock ``Experimental and theoretical evidence for the existence of absolute
  acoustic band gaps in two-dimensional solid phononic crystals.''\ {\em
  Physical Review Letters}, 86(14), 3012--3015.

\bibitem[\protect\citeauthoryear{}{Veres and Berer}{2012}]{veres2012complexity}
Veres, I.~A. and Berer, T. (2012).
\newblock ``Complexity of band structures: Semi-analytical finite element
  analysis of one-dimensional surface phononic crystals.''\ {\em Physical
  Review B}, 86(10), 104304.

\bibitem[\protect\citeauthoryear{}{Willis}{2016}]{willis2016negative}
Willis, J. (2016).
\newblock ``Negative refraction in a laminate.''\ {\em Journal of the Mechanics
  and Physics of Solids}, 97, 10--18.

\bibitem[\protect\citeauthoryear{}{Yang et~al.\@}{2002}]{yang2002ultrasound}
Yang, S., Page, J.~H., Liu, Z., Cowan, M.~L., Chan, C.~T., and Sheng, P.
  (2002).
\newblock ``Ultrasound tunneling through 3d phononic crystals.''\ {\em Physical
  review letters}, 88(10), 104301.

\bibitem[\protect\citeauthoryear{}{Yang et~al.\@}{2004}]{yang2004focusing}
Yang, S., Page, J.~H., Liu, Z., Cowan, M.~L., Chan, C.~T., and Sheng, P.
  (2004).
\newblock ``Focusing of sound in a 3d phononic crystal.''\ {\em Physical review
  letters}, 93(2), 024301.

\bibitem[\protect\citeauthoryear{}{Zen et~al.\@}{2014}]{zen2014engineering}
Zen, N., Puurtinen, T.~A., Isotalo, T.~J., Chaudhuri, S., and Maasilta, I.~J.
  (2014).
\newblock ``Engineering thermal conductance using a two-dimensional phononic
  crystal.''\ {\em Nature communications}, 5(3435).

\end{thebibliography}

\end{document}